\documentclass[aps, superscriptaddress, showpacs, floatfix, twocolumn]{revtex4-1}
\usepackage{amsmath}
\usepackage{hyperref}
\usepackage{bbm}
\usepackage{graphicx}
\usepackage{float}
\usepackage{multirow}
\usepackage[usenames,dvipsnames]{color}
\usepackage{soul}

\newcommand{\sro}{SrRuO$_3$\,}
\newcommand{\cro}{CaRuO$_3$\,}
\newcommand{\eV}{\mathrm{eV}}

\hypersetup{
  colorlinks = true, 
  urlcolor   = blue, 
  linkcolor  = blue, 
  citecolor  = blue  
}

\begin{document}
\title{Electronic correlations, magnetism and Hund's rule coupling\\ in the ruthenium perovskites SrRuO$_3$ and CaRuO$_3$}

\author{Hung T. Dang}
\affiliation{Institute for Theoretical Solid State Physics, JARA-FIT and JARA-HPC, RWTH Aachen University, 52056 Aachen, Germany}
\author{Jernej Mravlje}
\affiliation{Jo\v{z}ef Stefan Institute, Jamova 39, Ljubljana, Slovenia}
\author{Antoine Georges}
\affiliation{Coll{\`e}ge de France, 11 place Marcelin Berthelot, 75005 Paris, France}
\affiliation{Centre de Physique Th\'eorique, Ecole Polytechnique, CNRS, 91128 Palaiseau Cedex, France}
\affiliation{DQMC, Universit{\'e} de Gen{\`e}ve, 24 quai Ernest-Ansermet, 1211 Gen{\`e}ve 4, Switzerland}
\author{Andrew J. Millis}
\affiliation{Department of Physics, Columbia University, New York, New York 10027, USA}

\begin{abstract}

A comparative density functional plus dynamical mean field theory study of the pseudocubic ruthenate materials CaRuO$_3$ and SrRuO$_3$ is presented.  Phase diagrams are determined for both materials as a function of  Hubbard repulsion $U$ and Hund's rule coupling $J$. Metallic and insulating phases are found, as are ferromagnetic and paramagnetic states. The locations of the  relevant phase boundaries are determined.  Based on the computed phase diagrams, Mott-dominated and Hund's dominated regimes of strong correlation are distinguished. Comparison of calculated properties to experiments indicates that the actual materials are in the Hund's coupling dominated region of the phase diagram so can be characterized as Hund's metals, in common with other members of the ruthenate family.   Comparison of the phase diagrams for the two materials reveals the role played by
rotational and tilt (GdFeO$_3$-type) distortions of the ideal perovskite structure. The presence of magnetism in SrRuO$_3$ and its absence in CaRuO$_3$ despite the larger mass and larger tilt/rotational distortion amplitude of CaRuO$_3$  can be understood in terms of  density of states effects in the presence of strong Hund's coupling.  Comparison of the calculated  low-$T$ properties of CaRuO$_3$ to those of  SrRuO$_3$  provides insight into the effects of magnetic order on the properties of a Hund's metal. The study provides a simultaneous description of magnetism and correlations and explicates the roles played by band theory and Hubbard and Hund's interactions. 
\end{abstract}


\maketitle

\section{Introduction\label{sec:intro}}

The notion that the electronic properties of crystalline materials can to a large degree be understood in terms of the energy bands arising from the solution of the Schr\"odinger equation for a single electron in a periodic potential is fundamental to condensed matter physics and its applications. Electrons are charged and the inter-electron Coulomb interaction cannot be neglected.  Density functional theory (DFT), in essence a sophisticated mean field treatment of electron-electron interactions, provides a very good approximation to the interacting electron problem, enabling the theoretical description from first principles of many properties of many compounds. However DFT does not describe all electronic properties of all materials, and the cases where it fails can be taken to define the ``strong correlation problem''.

One dramatic example of strong correlations is the ``Mott'' insulator \cite{Mott49}: a material in which the correlations are so strong that they lead to insulating behavior in situations where band theory predicts a metal. Less extreme cases, where the interactions do not drive the material insulating but do lead to strong renormalization of electron velocity relative to band theory, to large and strongly temperature and frequency dependent electron lifetimes, and to  the occurrence of magnetic order, have been extensively documented \cite{Imada98}.

Transition metal oxides (TMOs)~\cite{Mott49,Imada98}  play a particularly important role in the investigation of electronic correlations. In many TMO materials the transition metal $d$ shells are partially filled. Interactions between electrons in the $d$ orbitals of a transition metal ion are characterized by a sizable effective Coulomb repulsion $U_\mathrm{eff}$ that is close in magnitude to the bandwidth $W$ of the $d$-derived bands and leads to the formation of spin and orbital degrees of freedom.  As a result the physics of TMO materials often differs sharply from the predictions of DFT and involves an intricate interplay of charge, spin and orbital degrees of freedom, which is furthermore highly sensitive to details of the crystal structure.

These issues have been intensively studied in the context of TMOs which crystallize in variants of the $AB$O$_3$ perovskite structure. In the ideal perovskite structure the $B$ site ions lie on the vertices of a simple cubic lattice; each $B$ site ion is octahedrally coordinated by oxygen.  Few members of the family of $AB$O$_3$ transition metal oxides crystallize in the ideal cubic structure: in most materials a mismatch between the size of the $A$ and $B$ site ions (``tolerance factor'' less than one) leads to a compressive strain on the $B$O$_3$ network. This strain is typically accommodated by a rotational and tilt (GdFeO$_3$-type) distortion of the $B$O$_6$ octahedra that diminishes the width of the $d$-derived bands and lowers the degeneracy of the $d$ multiplets.  

Particular attention has been given to  materials in which the $B$-site ion is drawn from the first transition metal row of the periodic table so that the $3d$ shell of the transition metal ion is partially occupied. In these materials the key physics is the correlation-induced metal-insulator transition (often referred to as the `Mott' transition \cite{Mott49} although actual atomic-scale physics may be more involved \cite{Zaanen85}). The prevailing understanding \cite{Fujimori92,Imada98} is that in most of these materials the basic ``correlation strength'' is related to the proximity of the material to the Mott transition (but see Refs.~\onlinecite{Mizokawa00,Park12,Johnston14,subedi_14} for the exceptional case of the nickelates) while the rotational and tilting distortions play a key role in determining this proximity.  For example, SrVO$_3$ (nominal valence $d^1$) crystallizes in the simple cubic structure and is a moderately correlated Fermi liquid \cite{Inoue98}. In CaVO$_3$, a small-amplitude GdFeO$_3$ distortion occurs; the material is still metallic but more correlated than SrVO$_3$ \cite{Inoue95,Inoue98,Makino98}. In LaTiO$_3$ the nominal $d$ valence is also $d^1$, however a larger-amplitude GdFeO$_3$ distortion is present and the material is a Mott insulator. In isoelectronic YTiO$_3$ the distortion amplitude and the insulating gap are larger than in LaTiO$_3$ \cite{Arima93,Imada98}. The differences between Sr- and CaVO$_3$ or between La- and YTiO$_3$ may be attributed to different amplitudes of the  GdFeO$_3$ distortion. Theoretical work~\cite{Pavarini04} showed that the key physics is a lifting of the degeneracy of the transition metal $t_{2g}$ levels; this is important  because the critical interaction strength needed to drive a Mott transition depends strongly on orbital degeneracy, see, e.g.  Ref. \onlinecite{Werner09} and references therein. (The differences between the V-based and Ti-based materials arise in part from difference in GdFeO$_3$ distortion amplitude and in part from the difference in relative electronegativities of Ti and V \cite{Dang14a,Dang14b}.)

However, proximity to a Mott insulating state is not the only cause of correlated electron behavior. In heavy-fermion materials a lattice version of the Kondo effect can lead to enormous mass enhancements and other exotic physics \cite{lohneysenRMP_07}. In transition metal oxides with nominal valences from $d^2$ to $d^8$ the Hund's coupling can play a crucial role in producing very large renormalizations even for materials far from a Mott transition \cite{Haule09,Werner08,Mravlje11,Medici11,yin11,Georges13}.

In this regard transition metal oxides where the transition metal is drawn from the $4d$ series are of particular interest. Because $4d$ orbitals have a much greater spatial extent than $3d$ orbitals, the effective bandwidth is larger and the $U_\mathrm{eff}$ is smaller,  suggesting that the $4d$ materials are in general likely to be farther from the Mott state than the $3d$ materials. Although many of the $4d$ series TMO   are indeed itinerant metals, signatures of strong correlations, such as enhancement of  the specific heat \cite{Bergemann03,Chubukov06}, magnetic transitions \cite{Mackenzie03} and evidence of other unusual electronic phases \cite{Grigera01}, are clearly present, in particular in the ruthenate family~\cite{Mackenzie03, Georges13}. Further, some members of the $4d$ series (for example Ca$_2$RuO$_4$) have been identified as Mott insulators \cite{Gorelov10}.  Thus in the $4d$-series transition metal oxides the issue of the  relative importance of Mott and Hund's correlations remains unclear, as does the role of the GdFeO$_3$ distortions.

Here, we explore these issues by focusing on the two of pseudocubic  ruthenates: \sro and \cro.  Both crystallize in GdFeO$_3$-distorted versions of the $AB$O$_3$ perovskite structure; with the distortion amplitude being larger in \cro than in \sro.  \sro is ferromagnetically ordered (a rather rare behaviour among $4d$ TMOs) below a Curie temperature $T_c\sim 160$\,K while \cro remains paramagnetic to lowest temperatures.  On the applied side, SrRuO$_3$ is a convenient electrode material, widely used as a substrate and magnetic ingredient in heterostructures and spin-valves~\cite{Bibes99,Hikita07,He11,koster12rmp}.  Basic scientific questions remain open, including their degree of correlation, the origin of the apparently non-Fermi-liquid properties evident in the optical spectra \cite{Kostic98,Lee02} and the reason for the magnetism, in particular why the apparently less strongly correlated material  \sro is magnetic while the apparently more correlated \cro is not. There is also fundamental interest in obtaining a better understanding of  ruthenates in general,  because insights gained in the study of the pseudocubic materials may shed light on   the unconventional superconductivity of Sr$_2$RuO$_4$~\cite{Mackenzie03} and the metamagnetism  and other phenomena observed in Sr$_3$Ru$_2$O$_7$~\cite{Grigera01}.

The question of the correct physical picture of the pseudocubic ruthenate perovskites (whether they should be regarded as weakly correlated itinerant metals or as strongly correlated systems) is the subject of controversy. On the experimental side,  photoemission experiments~\cite{Maiti05} do not detect Hubbard sidebands, suggesting that the materials are not in proximity to a Mott transition.  However, an earlier interpretation of the photoemission spectroscopy~\cite{fujioka97} indicated that  sizable renormalizations occur at low energies~\cite{kim05}. Optical spectroscopy~\cite{Kostic98,Dodge00,Lee02,Schneider14} indicates strong deviations from Fermi-liquid behavior, while transport experiments reveal very low Fermi liquid coherence scales ($7$\,K for SrRuO$_3$ \cite{Capogna02} and $1.5$\,K for CaRuO$_3$ \cite{Schneider14}) and large mass enhancements. 

On the theory side, early analyses \cite{Singh96,Mazin97} of the electronic structure based on spin-dependent density functional theory [local spin density approximation (LSDA) or spin-dependent generalized gradient approximation (GGA)]  correctly describe many of the magnetic properties. The ferromagnetism  in \sro was interpreted as the result of a Stoner instability, and the presence of magnetism in \sro and its absence in \cro was related to the Fermi-level density of states, which is  higher and more sharply peaked in \sro than in \cro. However the DFT calculations do not account for the low coherence scales and large mass renormalizations.    A more recent comparative study of magnetism using a range of band theoretic techniques including the density functional plus $U$ method concluded  that $U=0$ gives the  best description of  the experimentally observed transition temperatures \cite{Etz12}. Within LSDA, properties of SrRuO$_3$ and CaRuO$_3$ under strain were calculated~\cite{zayak06,zayak08} and the predicted occurrence of ferromagnetism in CaRuO$_3$ under tensile strain was recently observed~\cite{Tripathi2014}. On the other hand, many theoretical papers  including the  LSDA+$U$ work of Rondinelli and collaborators~\cite{rondinelli_08} and several dynamical mean-field theory (DMFT) investigations~\cite{laad08,Jakobi11,huang13,Granas14} assert that correlations beyond LSDA/GGA are important.

The existing literature thus suggests that the  challenge presented by the perovskite ruthenates is to develop a theory that includes the electronic correlations that provide the experimentally indicated mass enhancement and other renormalizations without spoiling the good account of the magnetic phase diagram  found in  density functional calculations.  In this paper we address this challenge by performing a systematic density functional plus dynamical mean field theory study that includes realistic electronic structure and investigates a wide range of potentially relevant interaction parameters. We calculate the phase diagram in the $(U,J)$  plane and by comparing calculated and measured properties we locate the perovskite  ruthenates in the  correlated Hund's metal region of the phase diagram. Effective masses are found to be large and coherence scales small in the paramagnetic  phase.  The greater tendency to magnetic ordering in \sro than in \cro is accounted for in a manner similar to that found in earlier electronic structure calculations. Our results also provide insight into the general issue of the effect of GdFeO$_3$ distortions on the effective correlation strength of Hund's metal. The low frequency properties are affected by the density of states (which in the ruthenates is reduced by GdFeO$_3$ distortions) while the global and higher frequency correlation strength is controlled by the inverse bandwidth (which is increased by GdFeO$_3$ distortions). 

The rest of this paper is organized as follows. Section~\ref{sec:formalism} describes the methods we use. Section~\ref{sec:phase_diagrams} presents our computed ferromagnetic/paramagnetic (FM/PM) and metal-insulator phase diagrams
and discusses the physics behind them. Section~\ref{sec:mass} discusses the differences of two ruthenates in the mass enhancement and the self-energy and uses this information to locate the materials on the  phase diagram of Sec.~\ref{sec:phase_diagrams}. The magnetic phase of SrRuO$_3$  is analyzed in detail in Sec.~\ref{sec:magnetic_sro}.  Section~\ref{sec:conclusions} presents a summary and prospects for future work. Appendices provide details of the calculations

\section{Crystal structure, electronic structure and model\label{sec:formalism}}

CaRuO$_3$ and SrRuO$_3$ crystallize in a $Pnma$ symmetry crystal structure related to the ideal cubic perovskite structure by a GdFeO$_3$ distortion corresponding to a tilt and rotation of each RuO$_6$ octahedron. The tilts and rotations alternate in a four-sublattice pattern. SrRuO$_3$ has a Ru-O-Ru bond angle of about $163^\circ$, in between the ideal perovskite Ru-O-Ru bond angle of $180^\circ$ and the Ru-O-Ru bond angle of $150^\circ$ observed in CaRuO$_3$  \cite{Jones89,Bensch90}.

Valence counting implies that in CaRuO$_3$ and SrRuO$_3$ the  Ru is in the $d^{4+}$ electron configuration with four electrons in the Ru $4d$ shell. The octahedral ligand field pushes the $e_g$ levels up in energy so  the relevant near-Fermi-surface bands are derived from Ru $t_{2g}$ symmetry $d$ states with some admixture of the O $2p$ states. Because Ru is a second-row transition metal ion, the $d$ states are expected to be more extended and the on-site interaction $U$ is expected to be weaker than for the first-row transition metal ions, indicating \cite{Georges13} that the materials are not in  the charge transfer regime of the Zaanen-Sawatzky-Allen phase diagram \cite{Zaanen85}. We therefore adopt the ``frontier orbital'' approach in which the low energy electronic properties are obtained from a multiband Hubbard model with hybridizations and level splittings obtained from the   near-Fermi-surface transition metal $d$-derived bands. 

The Hamiltonian takes the general form 
\begin{equation}\label{eq:Hamiltonian}
H = H_{\mathrm{kin}} + H_{\mathrm{onsite}},
\end{equation} 
where $H_{\mathrm{kin}}$ describes the dispersion of the bands derived from the frontier orbitals and $H_{\mathrm{onsite}}$ the additional interactions.  The chemical potential is set to ensure that these bands contain four electrons per Ru.

To define the near-Fermi-surface bands of  $H_{\mathrm{kin}}$ precisely we use the non-spin-polarized generalized gradient approximation (GGA) as implemented in the {\sc Quantum ESPRESSO} density functional code \cite{QE-2009,QEPseudo} to obtain electronic band structures and then project the resulting bands onto maximally-localized Wannier functions (MLWF) \cite{Marzari97,Souza01} using the {\sc wannier90} code \cite{Mostofi08} (details are given in Appendix~\ref{Appendix:BandTheory}). 

In most of this paper we construct  $H_{\mathrm{kin}}$ by projecting  the Kohn-Sham Hamiltonian onto $t_{2g}$-symmetry Wannier functions centered on the Ru sites. This procedure captures correctly all of the electronically active frontier orbitals and provides a reliable  description of the phase diagram and quasiparticle properties. However, as will be discussed in detail in  Sec.~\ref{sec:magnetic_sro}, this procedure leads to an overestimate of  the magnetic moment in the magnetically ordered phase. Obtaining a correct estimate of the ordered moment requires inclusion of bands derived from Ru $e_g$ states. In our analysis of the magnetic state the $e_g$-derived bands are therefore retained, but because the $e_g$-derived states  are far from the Fermi level, they are treated by the mean-field approximation used in Ref. \onlinecite{Held07}.

The $t_{2g}$ orbitals are treated dynamically. As usual in studies of transition metal oxides, the interaction Hamiltonian  is taken to be site-local and to have  the rotationally invariant Slater-Kanamori form \cite{Imada98}. We use the form appropriate \cite{Georges13} for intra-$t_{2g}$ orbitals, since these are the primary focus of this work
\begin{equation}\label{eq:onsite_SlaterKanamori}
\begin{split}
H_{onsite} & = U\sum_{\alpha}n_{\alpha\uparrow}n_{\alpha\downarrow}  + (U-2J)\sum_{\alpha\neq\beta} n_{\alpha\uparrow}n_{\beta\downarrow} + \\
& + (U-3J)\sum_{\alpha > \beta,\sigma}n_{\alpha\sigma}n_{\beta\sigma} + \\
& + J\sum_{\alpha\neq\beta} ( c^\dagger_{\alpha\uparrow}c^\dagger_{\beta\downarrow}c_{\alpha\downarrow}c_{\beta\uparrow}
+ c^\dagger_{\alpha\uparrow}c^\dagger_{\alpha\downarrow}c_{\beta\downarrow}c_{\beta\uparrow}
),
\end{split}
\end{equation}
where $ \alpha,\beta $ are orbital indices and $ \sigma $ is the spin index. Different values of $U$ and $J$ have been used for ruthenates in the literature. For Sr$_2$RuO$_4$, constrained LDA~\cite{Pchelkina07} gives $(U,J)$ values of $(3.1~\eV,0.7~\eV)$ while the values  $(2.3~\eV,0.25~\eV)$~\cite{Mravlje11} and $(2.6~\eV, 0.26~\eV)$ \cite{Vaugier12} have been obtained from  constrained random phase approximation (cRPA) method. As discussed above, for the perovskite ruthenates $U$ values ranging from zero to rather large numbers have been employed. For this reason and because the behavior of the model for general parameters is of theoretical interest, we consider a range of values for $U$ and $J$ in this paper. However we restrict attention to the regime $U>3J$ where the effective on-site interaction is positive in all channels.

To treat the onsite interaction [Eq.~\eqref{eq:onsite_SlaterKanamori}], we employ single-site dynamical mean-field theory (DMFT) \cite{Georges96}. This method allows us to map the Hamiltonian [Eq.~\eqref{eq:Hamiltonian}] into a multiorbital impurity model embedded in a fermion bath. The impurity model is solved by using the hybridization expansion version of the continuous time quantum Monte Carlo (CT-HYB) \cite{Werner06} implemented in the TRIQS code \cite{triqsproject} for rotationally invariant interaction using  conserved quantities \cite{Parragh12}  to speed up the calculations. 

Care is required in the definition of the impurity model. Each Ru ion is at the center of an octahedron defined by six oxygen ions.  The $Pnma$ structure means that the local symmetry axes of a given RuO$_6$ octahedron are not parallel to the axes that define the global crystal structure. If the $e_g$ and $t_{2g}$ combinations of the $d$-derived states in a single octahedron are defined with respect to the global, rather than the local, symmetry axes, the impurity model will contain off-diagonal terms which mix the different orbitals at the single-particle level. This causes a severe sign problem for the CT-HYB solver \cite{Gull11}. It is preferable to avoid this complication, following Ref.~\onlinecite{Dang13}, by using a local basis with symmetry axes aligned along the octahedral directions appropriate to a given Ru, instead of the global axes.  When restricted to the $t_{2g}$ manifold only, the MLWF approach used here produces orbitals that are already  aligned with respect to the local axes of the appropriate octahedron so the hybridization function is essentially diagonal. Thus  if only $t_{2g}$ orbitals are retained  the DMFT calculation is  straightforward:  the impurity model with hybridization function defined directly from the projection of the Kohn-Sham Hamiltonian onto the Wannier states is   solved  for one Ru site. The  self-energies for the other Ru sites are then constructed by applying appropriate rotation operators.   If the MLWF procedure is applied to the full $d$ manifold (including both $t_{2g}$ and  $e_g$ orbitals) then the resulting orbitals turn out  not to be  aligned to the local symmetry axes and an additional change of basis is required before solving the impurity model (see Appendix~\ref{Appendix:BandTheory}). From the solution of the dynamical mean field equations we determine the phase (metal versus insulator, paramagnetic versus ferromagnetic) and some properties of the phases, in particular the quasiparticle mass enhancement and the magnetic moment. Details of our procedure for determining different phases are given in Appendix~\ref{sec:criteria}.

\section{Qualitative Physics\label{sec:phase_diagrams}} 

\subsection{Electronic structure: Density of states}

\begin{figure}[htbp]
  \centering
  \includegraphics[width=\columnwidth]{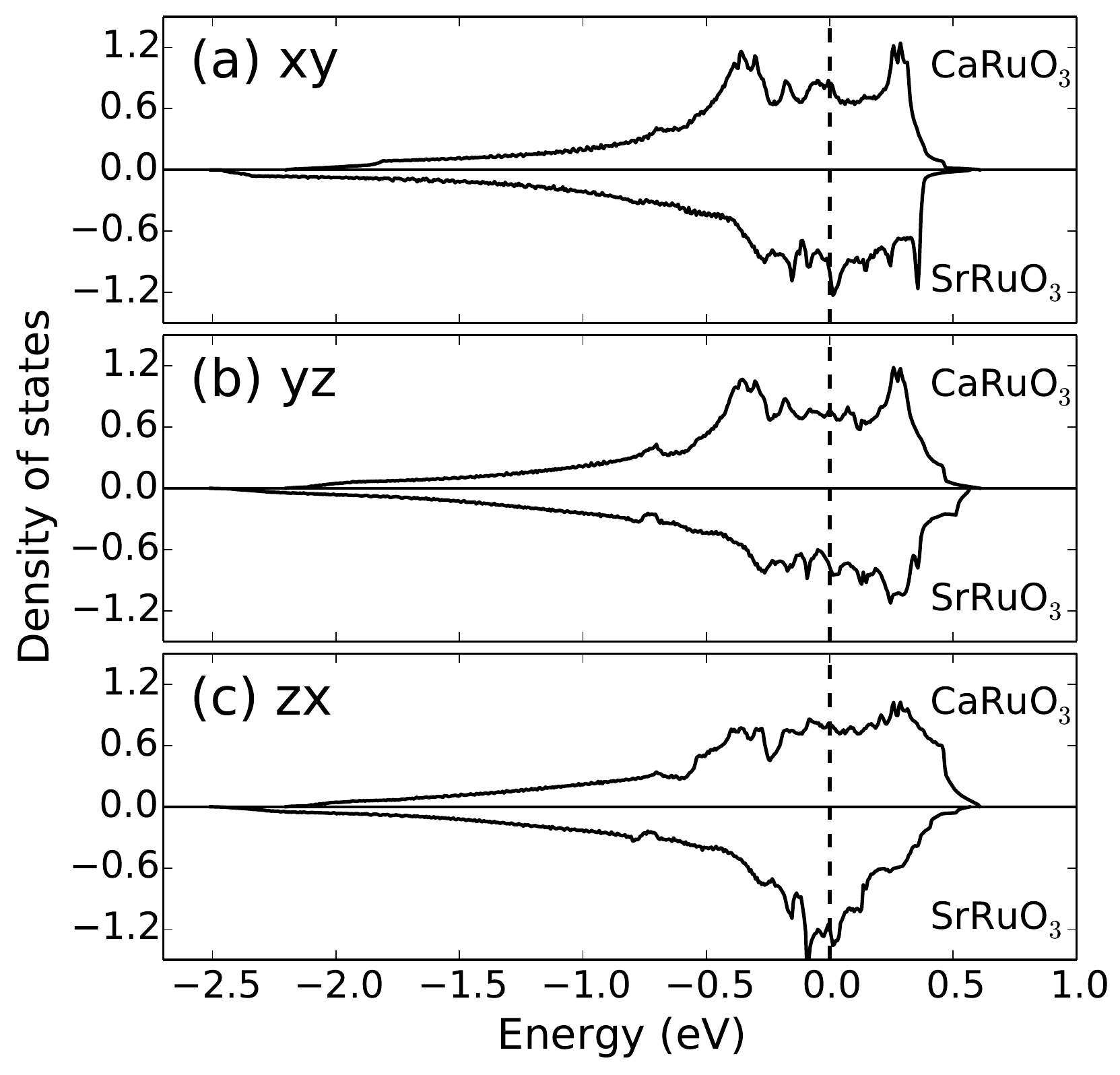}
  \caption{\label{fig:dft_mlwf_dos} Orbitally-resolved density of states generated from MLWF fittings of the DFT calculations for CaRuO$_3$ (SrRuO$_3$) on the positive (negative) half plane. Each panel is corresponding to each of the $t_{2g}$ orbitals. The $(xy, yz, zx)$ occupancies are $(1.40,1.36,1.23)$ for \cro, and $(1.34,1.26,1.40)$ for \sro, respectively. In the same order, the DOS at the Fermi level $\nu_F$ (states/eV per Ru atom) for each of the three $t_{2g}$ orbitals are $(0.83,0.73,0.79)$ for \cro, and $(1.00,0.77,1.19)$ for \sro. The vertical dashed line marks the Fermi level.}
\end{figure}

Figure~\ref{fig:dft_mlwf_dos} shows the orbitally-resolved near-Fermi-surface density of states (DOS). The  DOS  of the two materials are similar, as expected from the essentially identical quantum chemistry, but the difference in the magnitude of the GdFeO$_3$ distortion occurring in the two materials leads to  two important differences in the DOS.   First, the $t_{2g}$-derived bands in SrRuO$_3$ are approximately $10\%$ wider than those of CaRuO$_3$ (\sro bandwidth $\approx 3.0$\,eV, as compared to $\approx 2.7$\,eV for CaRuO$_3$). To the extent that correlation effects scale as the ratio of an interaction strength to a bandwidth, this would suggest that \cro would be the more strongly correlated material. 

However, \sro has the larger density of states at the Fermi level. This can be traced back to the van Hove singularity  of the undistorted cubic structure, which happens to lie very close to the Fermi level. Because the GdFeO$_3$ distortion lifts the degeneracy of the $t_{2g}$ levels, it splits the van Hove peak into three features. In \sro the splitting is small and the density of states remains large. In \cro the splitting is larger, leading to a smaller Fermi-level DOS. To the extent that correlation effects are related to the Fermi-level density of states, this suggests that \sro would be the more strongly correlated material. In particular, the Stoner model of ferromagnetism relates the presence of magnetic order to the value of the product of an interaction constant and the Fermi-level density of states \cite{Stoner27}, so the density of states difference would suggest (in agreement with experiment and with the DFT work of Refs.~\onlinecite{Allen96,Singh96,Mazin97}) that \sro is more likely to be magnetic than \cro. Further, particularly in \cro,  the splitting creates a  density of states peak below the Fermi level. The considerations of Ref.~\onlinecite{Dang13} building on previous work of Vollhardt and collaborators \cite{Vollhardt97,Ulmke98,Wahle98,Held98,Chan09} suggest that this peak is unfavorable to magnetism.

\subsection{Phase diagrams}

\begin{figure}[t]
  \centering
  \includegraphics[width=0.9\columnwidth]{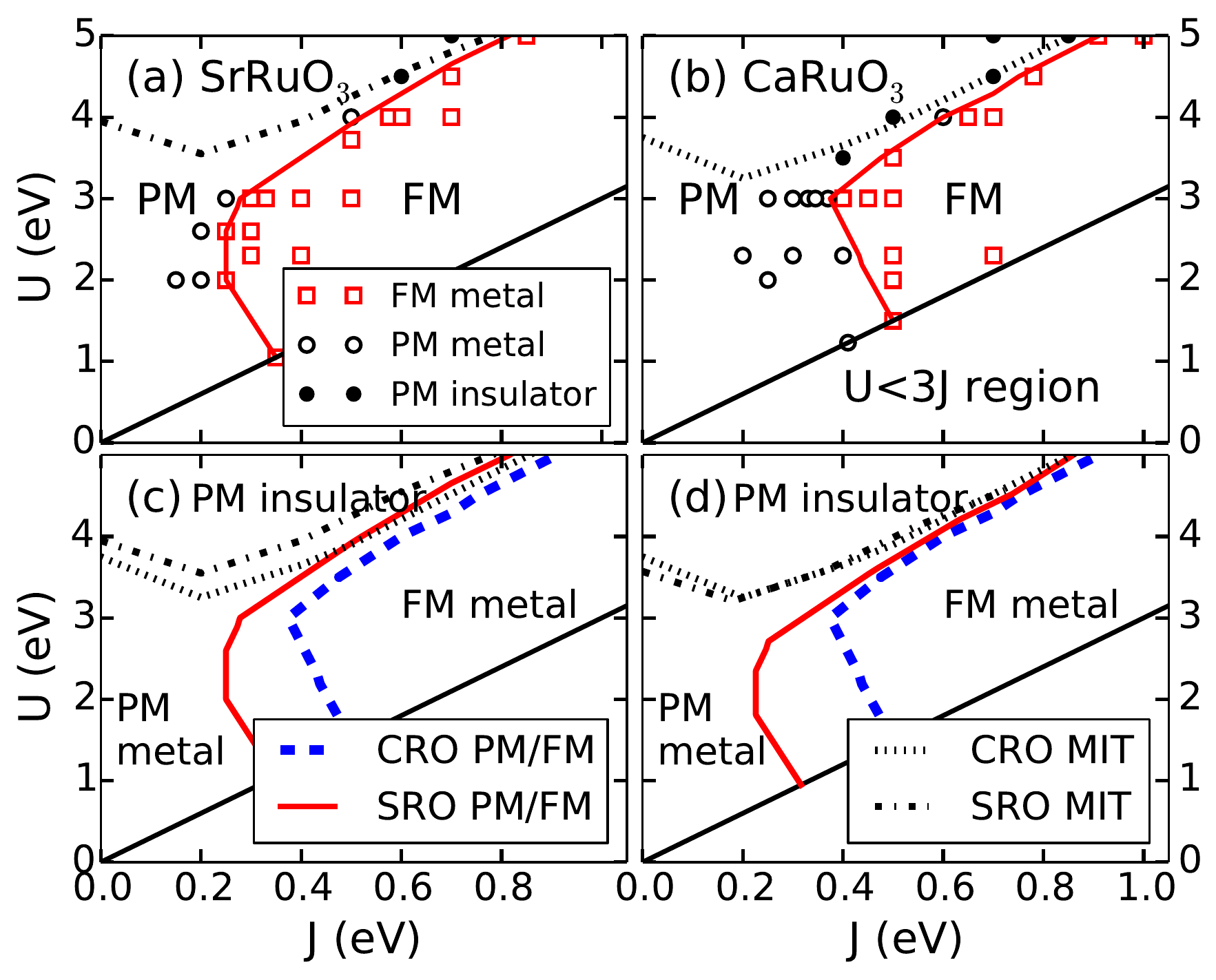}
  \caption{\label{fig:phase_diagram}(Color online) Ferromagnetic metal/paramagnetic metal (FM/PM) and metal/Mott insulator (MIT) phase diagrams. The  dotted lines indicate the metal-insulator phase boundary, with the region above the lines being insulating. The heavy solid line (red on-line) indicates the boundary of the ferromagnetic region, while the light line (black on-line) separates the physically relevant positive effective interaction ($U>3J$) from the unphysical negative interaction ($U<3J$) region.  
  (a,b) Phase diagrams  for SrRuO$_3$ (a) and CaRuO$_3$ (b) obtained from the DFT+DMFT procedure described in the text. Open circles (black on-line) indicate $(U,J)$ points for which properties were computed and a paramagnetic metallic state was found. Closed circles (black on-line) indicate paramagnetic insulating solutions.  Open squares (red on-line)  indicate $(U,J)$ points for which a ferromagnetic metallic state was found.   (c) The phase diagrams for SrRuO$_3$ and CaRuO$_3$ plotted together. Ferromagnetic phase boundary of CaRuO$_3$ indicated by heavy dashed line (blue on-line).   (d) Phase diagrams for SrRuO$_3$ and CaRuO$_3$ plotted together, but with the phase boundary of SrRuO$_3$ rescaled by the ratio of the SrRuO$_3$ bandwidth to the CaRuO$_3$ bandwidth. Note that there is no FM insulating phase found in these phase diagrams.}
\end{figure}

Figure~\ref{fig:phase_diagram} displays the phase diagrams in the plane of the Hubbard $U$ and the Hund's coupling $J$, determined by the procedure described in Appendix~\ref{sec:criteria}. We considered metallic and insulating phases, paramagnetic and ferromagnetic phases. Antiferromagnetism was not studied.

Focus first on the upper right panel, which presents results for CaRuO$_3$. We see that as the interaction strength $U$ is increased at fixed $J$, there is a phase transition from a metal (which may be paramagnetic or ferromagnetic) to a Mott insulator. As $J$ is increased at fixed $U$, a transition to a ferromagnetic metal phase occurs but no ferromagnetism was found in insulating phases in the $U$ and $J$ ranges that we studied. From this phase diagram one can identify two regimes: at large values of $U$, near the metal-insulator phase boundary, properties are most sensitive to the value of the effective Hubbard interaction. Note in particular that at large $J$ the phase boundary becomes a straight line with slope $U-3J$. The quantity $U-3J$ is the effective Hubbard interaction (correlation strength) relevant to the Mott transition because it gives the lowest  energy cost for a valence change from $2d^4$ to high spin $d^3d^5$. In contrast, far from the metal-insulator phase boundary, the Hund's coupling $J$ is the key parameter: by increasing $J$ a ferromagnetic phase is induced and there is a  range of $U$ in which the location of the critical boundary is only weakly dependent on $U$. 

The upper left panel displays our results for SrRuO$_3$.  The same phases are found but the difference in GdFeO$_3$ distortion amplitude causes  the location of the phase boundaries to be different. To highlight  the differences between the two calculations  we present in Fig.~\ref{fig:phase_diagram}(c) a superposition of the two phase diagrams.  At larger $U$ the phase boundaries are parallel (and controlled by $U-3J$), with \cro requiring a slightly smaller value of $U$ to be driven into the Mott phase, as expected from the smaller bandwidth and larger $t_{2g}$-level splitting following from the  larger distortion in \cro. However, in this region of the phase diagram the magnetic phase boundaries are very similar, and an extreme fine-tuning of $(U,J)$ would be required to  account for the fact that \cro is a paramagnetic metal, while \sro is a ferromagnetic metal.  

A much more significant difference between phase boundaries for the two structures is found at smaller values of $U\lesssim 3$~eV. There, the CaRuO$_3$ phase diagram has a significantly smaller region of ferromagnetism than the SrRuO$_3$ one. Thus, in this regime much less fine-tuning of the parameters is needed  to correctly account for the difference in magnetic behavior of the two materials. As discussed in Sec.~\ref{sec:mass}, there are also other experimental indications suggesting that these two materials should be viewed as being in this Hund's coupling dominated regime. We also note that the cRPA values of $U,J$ found for the related Sr$_2$RuO$_4$ material  \cite{Vaugier12}  are in the moderate $U$, larger $J$ region where \cro is paramagnetic but \sro is ferromagnetic. 

We now discuss further the qualitative origin of the observed differences between the phase diagrams of the two structures, by presenting in Fig.~\ref{fig:phase_diagram}(d)  a superposition of the phase diagrams of the two materials, but with the $U$ and $J$ values for \sro rescaled by the ratio $r=1.11$ of the \sro to the \cro bandwidth. At larger $U$ the phase diagrams for both the metal-insulator transition  and magnetism coincide in the rescaled plot.  This indicates that in the Mott-dominated region the physics is controlled by the ratio of the interaction strengths to the bandwidths and depends only weakly on for example the Fermi-level DOS. However, in the smaller $U$ regime, the magnetic phase diagrams do not coincide, indicating that in this regime the physics is clearly not controlled solely by the difference in bandwidths. 

Instead, the substantial difference in the critical $J$ required to drive the ferromagnetic transition is associated to the DOS in the near-Fermi-level region.  One important property is the value of the DOS at the  Fermi level $\nu_F$. In the standard Stoner theory \cite{Stoner27} magnetism is associated with a value greater than unity of a dimensionless interaction parameter $I$  constructed as the product of an appropriate interaction energy and the Fermi level density of states ($I=U \nu_F > 1$). Clearly the larger DOS in \sro makes it easier for the Stoner parameter to exceed the critical value and as discussed by Mazin and collaborators \cite{Singh96,Mazin97} spin dependent DFT calculations indeed indicate a Stoner parameter slightly greater than unity for \sro and slightly less for \cro. It is also worth noting that the Stoner theory is in essence a Hartree approximation. When the correlation is fully treated, other factors such as the energy derivative  of the DOS at the Fermi level and indeed the structure of the DOS far from the Fermi level are also important and provide significant corrections to the simple Stoner estimate. Dynamical mean field studies of related models \cite{Vollhardt97,Ulmke98,Wahle98,Dang13} indicate that for systems with carrier concentration such that the $d$ shells are less than half occupied such as La$_{1-x}$Sr$_x$VO$_3$ ($d$ valence $d^{2-x}$) ferromagnetism is favored if the DOS peak is at or below the Fermi level \cite{Dang13}. A particle-hole transformation allows us to relate the results of Ref.~\onlinecite{Dang13} (derived for a system with valence  near $d^2$) to the ruthenates (valence $d^4$), concluding that in the ruthenates a DOS peak at or above the Fermi level favors ferromagnetism. Therefore, as seen in Fig.~\ref{fig:dft_mlwf_dos}, the larger distortion of CaRuO$_3$ produces below-Fermi-level density of states peaks, thus disfavoring ferromagnetism, while in  SrRuO$_3$  the DOS peaks concentrate at the Fermi level and ferromagnetism is favored.

\section{Self-Energy, Mass Enhancement and Quasiparticle Lifetime\label{sec:mass}}

In this section we study the electron self-energy and quasiparticle properties, choosing interaction parameters $U=2.3$ and $3$~eV as representatives of the Hund's metal and Mott-dominated regimes respectively. These values are  similar to those obtained for the related material Sr$_2$RuO$_4$ from {\it ab initio} estimations using  constrained DFT \cite{Pchelkina07} and constrained RPA \cite{Mravlje11} approaches, respectively.  We fix $J=0.4$~eV as a representative value for which the Ca material is paramagnetic and the Sr material is ferromagnetic.  
\begin{figure}[t]
  \centering
  \includegraphics[width=\columnwidth]{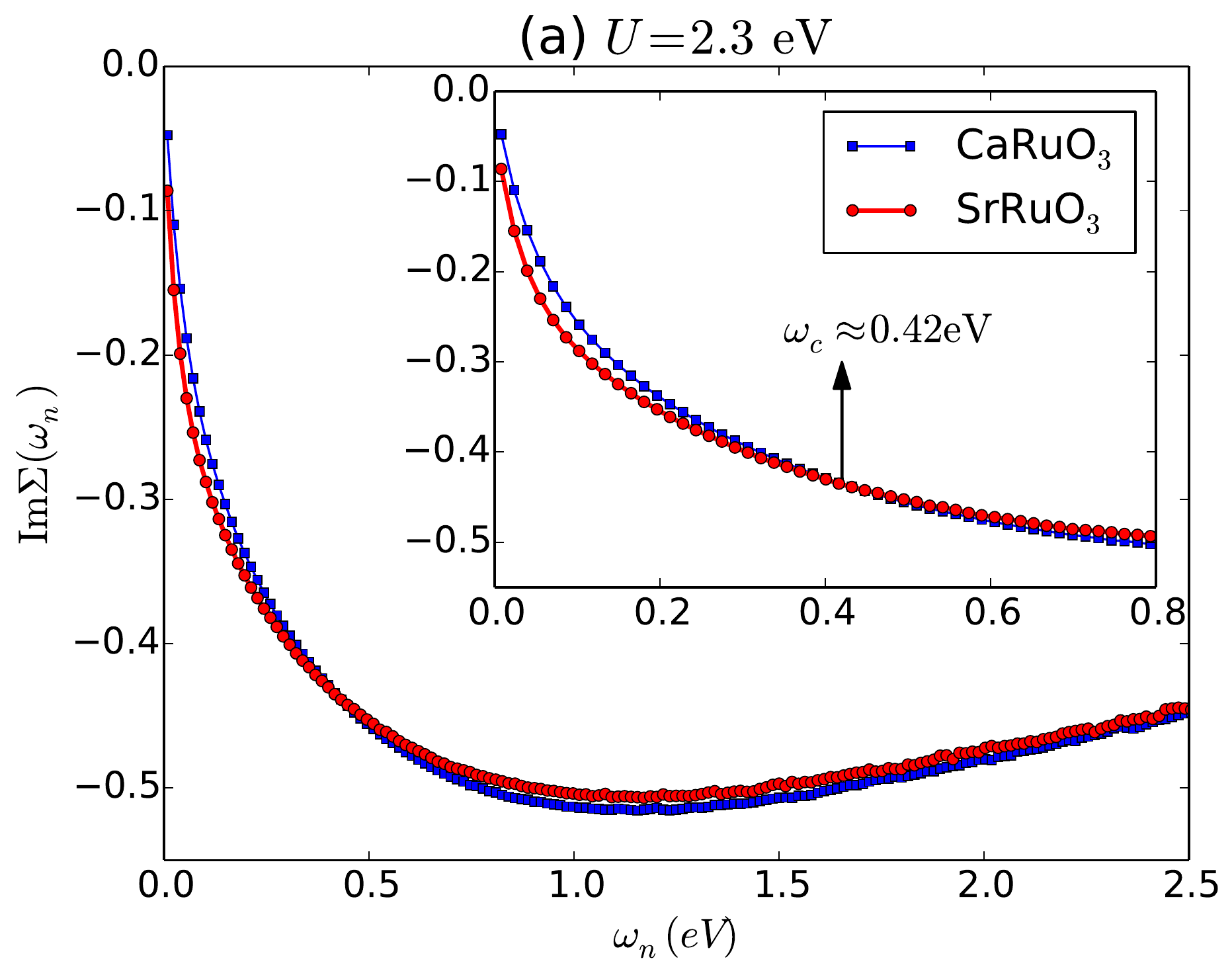}
  \includegraphics[width=\columnwidth]{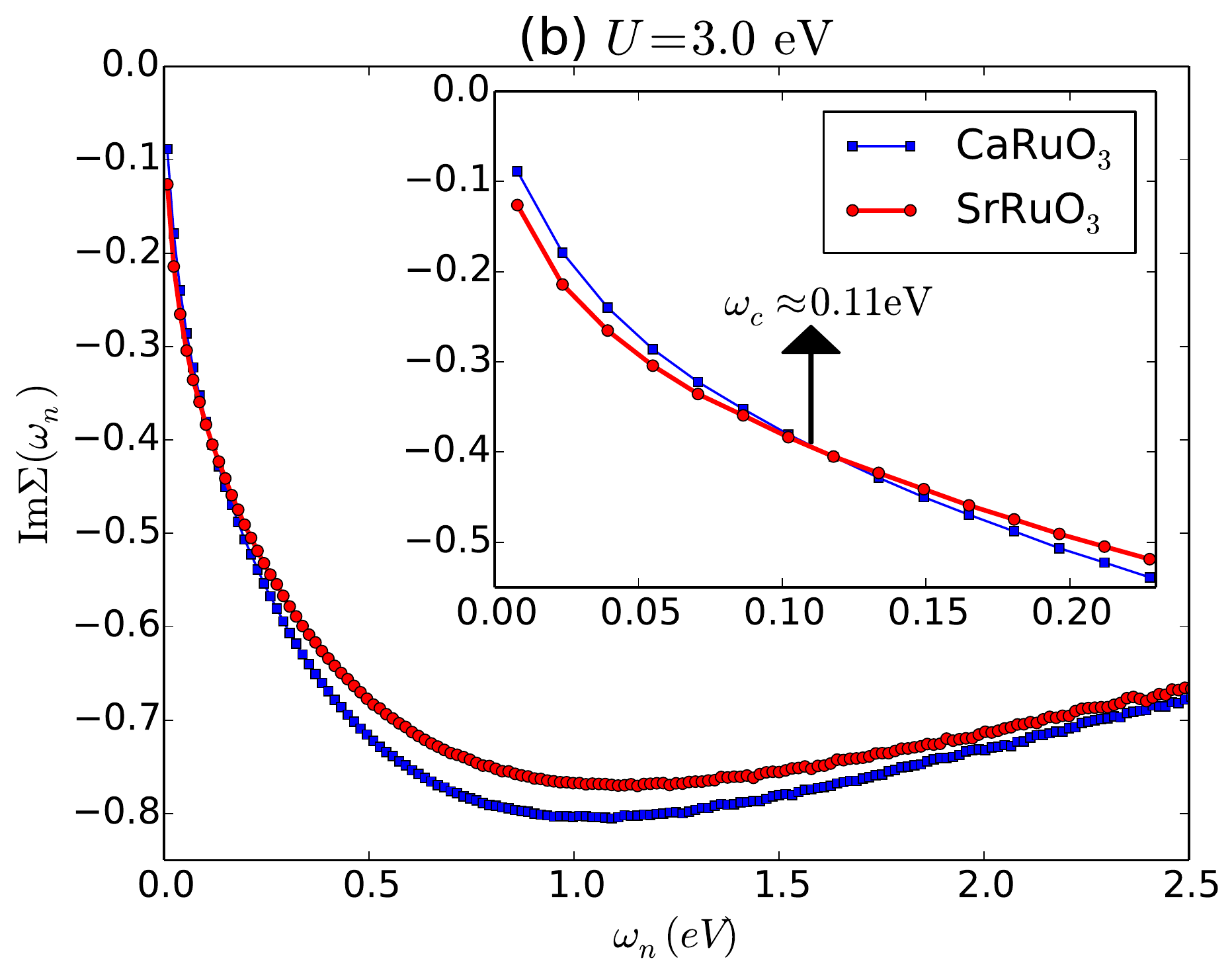}
  \caption{\label{fig:self_energy}(Color online) The imaginary part of the Matsubara self-energy $\mathrm{Im}\Sigma(i\omega_n)$ averaged over the three $t_{2g}$ orbitals,  calculated at $U=2.3$ (a) and $3$\,eV (b) with $J=0.4$\,eV, $T=0.0025$\,eV  for \sro and \cro. The inset provides an expanded version of the low frequency regime, allowing   the crossing point $\omega_c$ to be distinguished. Calculations are restricted to the paramagnetic order.}
\end{figure}

Figure~\ref{fig:self_energy} displays the imaginary parts of the  computed Matsubara-axis DMFT self-energies restricted to the paramagnetic phase over a wide range of Matsubara frequencies. We see that at low frequencies the self-energy of the Sr-compound is larger, indicating that for quantities dominated  by low energies the Sr material may be viewed as more correlated.  On the other hand, above a frequency $\omega_c$  the self-energy of the Ca material is larger,  reflecting the effect of the difference of bandwidths on the effective correlation strength at higher energies. The Hubbard $U$ and Hund's coupling $J$ compete in this respect. At smaller $U$ [Fig.~\ref{fig:self_energy}(a)], the Hund's coupling effect is stronger,  resulting in a wider range of low frequency (larger $\omega_c \approx 0.4$~eV) in which SrRuO$_3$ is more correlated. For the larger  $U$ close to the MIT phase boundary, Mott physics associated with $U$ becomes stronger,  as signaled by the decrease of $\omega_c$ by a factor of $\sim 4$ [$\omega_c\approx 0.1$~eV in Figure~\ref{fig:self_energy}(b)].

\begin{figure}[t]
    \centering
    \includegraphics[width=0.9\columnwidth]{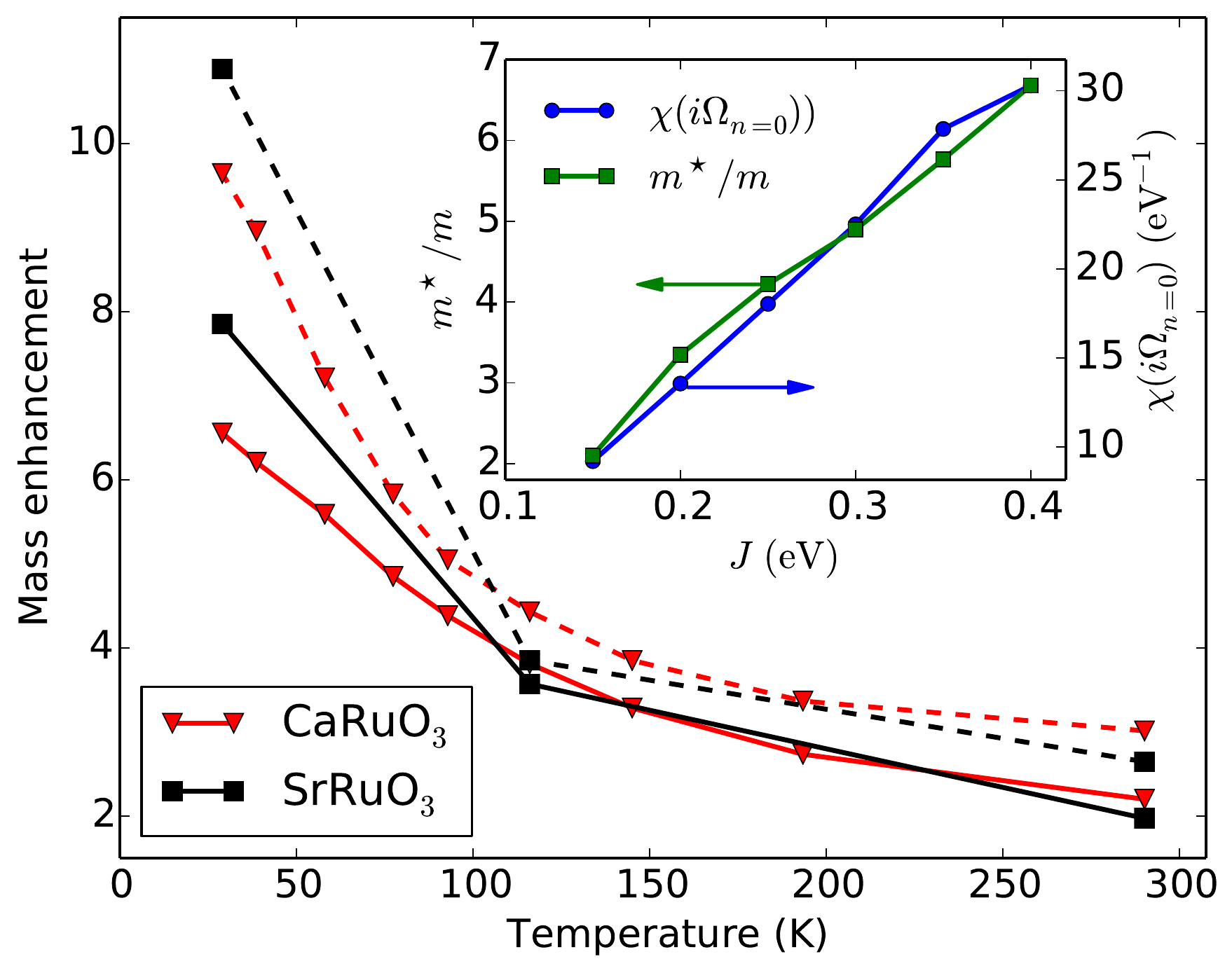}
    \caption{\label{fig:mass_enhancement}(Color online) 
The mass enhancement $Z^{-1}$ computed as described in the text for SrRuO$_3$ (squares, black on-line) and CaRuO$_3$ (triangles, red on-line) at $J=0.4$\,eV and $U=2.3$\,eV (solid lines) and $3$\,eV (dashed lines)  as a function of temperature.  In these calculations magnetism is suppressed: all results are for the paramagnetic phase of the model. Inset: comparison of mass enhancement  (squares, green on-line, left $y$ axis) and zero Matsubara frequency spin-spin correlation function $\chi(i\Omega_{n=0})=\int_0^\beta \langle S_z(\tau)S_z(0)\rangle d\tau$  with $S_z = \frac{1}{2} (N^{tot}_\uparrow - N^{tot}_\downarrow)$ (circles, blue on-line, right $y$ axis) computed  for CaRuO$_3$ at temperature $T=2.5~\mathrm{meV}\approx29$\,K with $U=2.3$\,eV and plotted as a function of $J$.}
\end{figure}

We now turn to two key physical quantities characterizing quasiparticles in the metallic state, namely the  effective mass enhancement $m^\star/m$ (directly related in the local DMFT approach to the quasiparticle weight $Z$)  and the quasiparticle scattering rate (inverse lifetime)   $\Gamma$.  These are defined from the real and imaginary part of the retarded self-energy on the real-frequency axis by: 
\begin{eqnarray}
\label{eq:mass_enhancement}
\frac{m^{*}}{m}&=&\frac{1}{Z} =  
1 - \frac{\partial}{\partial\omega}\mathrm{Re}\Sigma(\omega+i0^+)|_{\omega=0},\\ \nonumber
\Gamma\,&=&\,-Z\mathrm{Im}\Sigma(\omega+i0^+)|_{\omega=0}.
\end{eqnarray}
Inferring  real-axis quantities from Matsubara-frequency data in general requires analytical continuation. If, however, the low frequency properties are reasonably well described by the Fermi liquid fixed point  (as is the case for the parameters we study),  the low frequency limit of the  real frequency self-energy may be inferred with reasonable accuracy from the data at small Matsubara frequencies, with $1-Z^{-1}\approx d\mathrm{Im}\Sigma(i\omega_n)/d\omega_n|_{\omega_n\rightarrow 0}$ and $Z^{-1}\Gamma=-\mathrm{Im}\Sigma(i\omega_n\rightarrow 0)$.  In practice, we extract $Z$ and $\Gamma$ by  fitting a fourth-order polynomial to the first six Matsubara-axis data points for $\mathrm{Im}\Sigma(i\omega_n)$  and computing the needed quantities from the fitting function. 

Figure~\ref{fig:mass_enhancement} shows the estimated mass enhancement  for the two materials  at the two values of $U$ under consideration. The calculations are restricted to the paramagnetic state and show a strong temperature dependence, which is a manifestation of the low quasiparticle coherence scale associated with the formation and slow fluctuation of a local moment with $S \gg 1/2$ (in other words, with Hund's metal physics) \cite{Werner08,Georges13}.    As will be seen below, in the ferromagnetic state of SrRuO$_3$ the temperature dependence is cut off  because the ferromagnetic order quenches the slow spin fluctuations. To reinforce this point we show in the inset that the mass enhancement and zero Matsubara frequency impurity model spin correlation function have identical $J$ dependence.   Comparison to Fig.~\ref{fig:dft_mlwf_dos} shows that the local spin fluctuations involve an energy scale that is much lower than that of the bare density of states. 

\begin{figure}[ht]
 \centering
 \includegraphics[width=0.95\columnwidth]{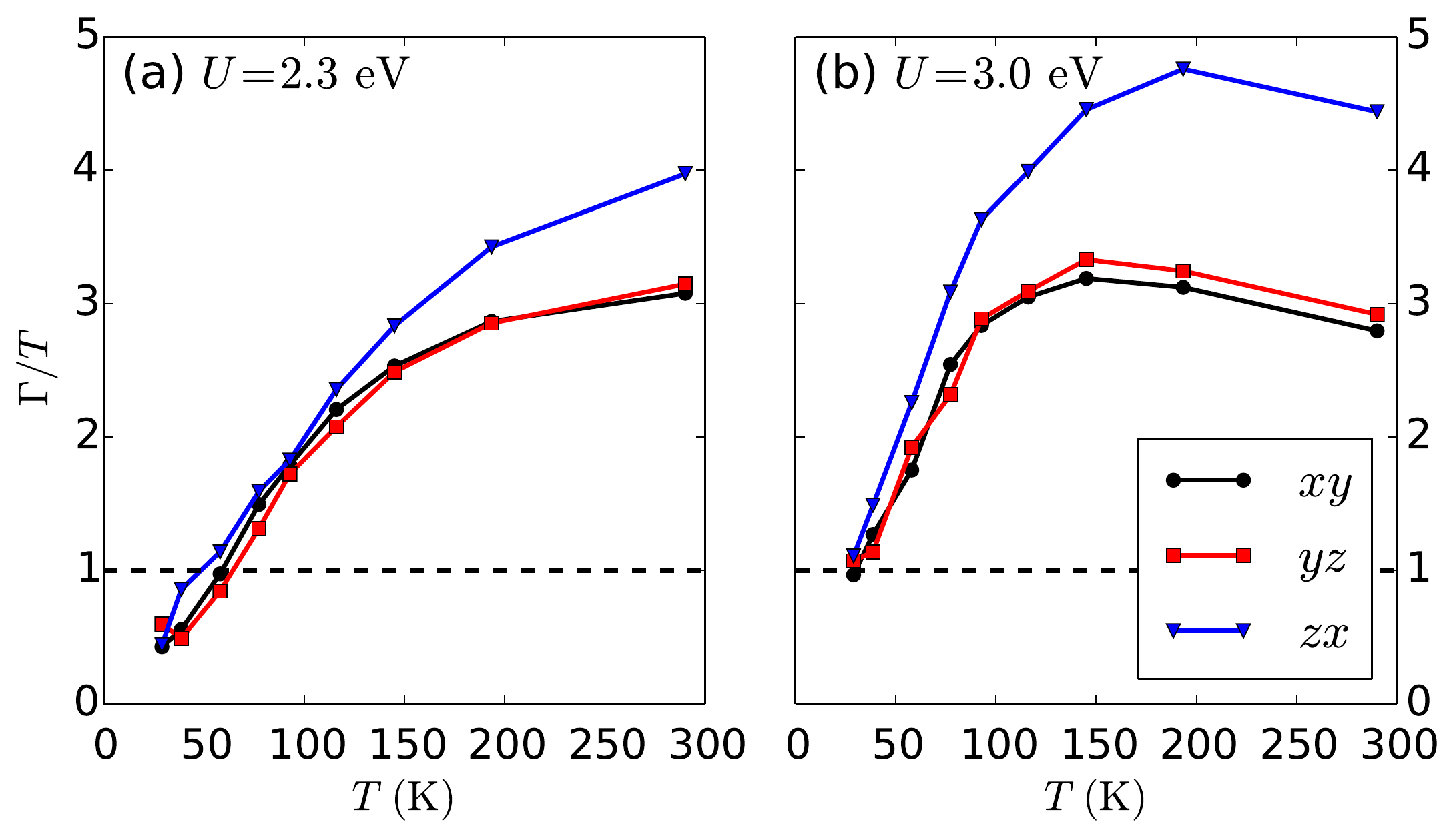}
 \caption{\label{fig:coherence}(Color online) Inverse of the quasiparticle lifetime, $\Gamma$ (Eq.~\eqref{eq:mass_enhancement}) calculated for CaRuO$_3$ at $J=0.4$\,eV and $U$-values indicated, divided by temperature and plotted as a function of  temperature, for each of the $t_{2g}$ orbitals. Linearity corresponds to Fermi liquid behaviour $\Gamma\propto T^2$. The horizontal dashed line marks the boundary ($\Gamma \le T$)  of the coherence region. The three orbitals differ slightly in occupancy; the closer the orbital is to half filling the more strongly it is correlated. \cite{Mravlje11}. The convention for the orbitals is given in Appendix~\ref{Appendix:BandTheory}.}
\end{figure}

In our simulations we were unable to reach temperatures low enough to observe the saturation of the mass to its $T\rightarrow 0$ limit in the paramagnetic phase. Nevertheless, we find the quasiparticles are becoming well defined at the lowest temperatures reached in our simulations. Figure~\ref{fig:coherence}  presents the temperature dependence of $\Gamma/T$ calculated for CaRuO$_3$ at $U=2.3$ and $3$\,eV with $J=0.4$\,eV.  Below $T=70$\,K for $U=2.3$\,eV  and below $T= 30$\,K for $U=3$\,eV  the scattering rate becomes smaller than temperature, which is indicative of coherence. We expect that as the temperature is lowered further below the coherence scale the mass will saturate. From these considerations we estimate the $T=0$ mass enhancements to be about 7 and $11$ for the $U=2.3$ and $3$\,eV, respectively.

We remark that Fig.~\ref{fig:coherence} implies that the quasiparticle scattering rate $\Gamma$ varies as $T^2$ up to $T\approx 150$\,K, even though the characteristic Fermi liquid signatures in physical observables (temperature independent mass/specific heat coefficient and magnetic susceptibility along with quadratic transport scattering rate) are only evident below much lower temperatures (lower than the lowest temperature accessible in our $U=2.3$, $J=0.4$\,eV calculations). This behavior qualifies CaRuO$_3$ as a `hidden Fermi liquid' \cite{xu_13,deng_14} in which although the temperature dependence of e.g. the resistivity deviates from $T^2$ above a very low temperature, the quasiparticle scattering rate remains $\sim T^2$ up to much higher temperatures and  the deviation from the Fermi liquid temperature dependence expected for the resistivity is attributable to a temperature dependence of the quasiparticle weight. 

\begin{figure}[t]
    \centering
    \includegraphics[width=0.8\columnwidth]{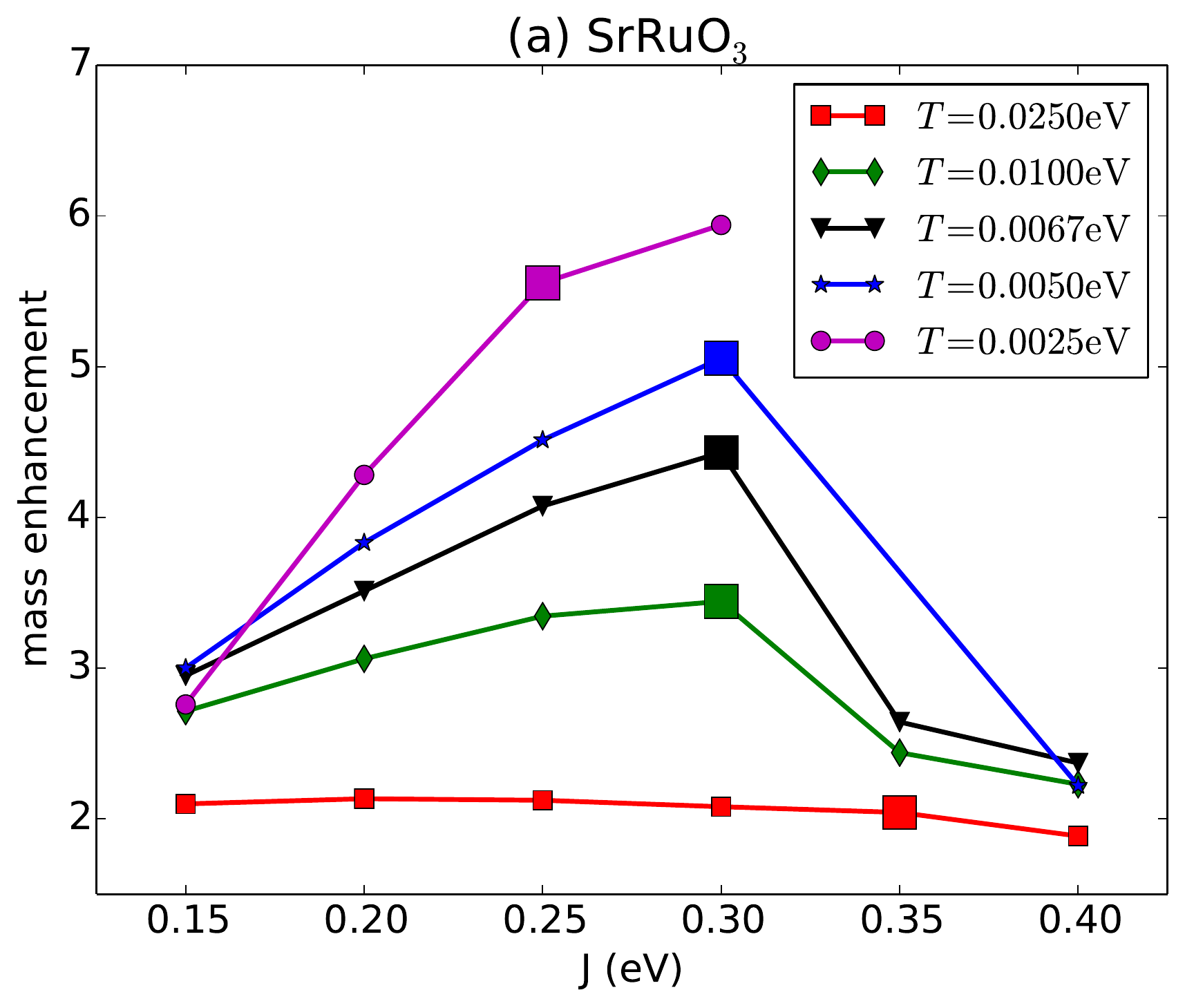}
    \includegraphics[width=0.8\columnwidth]{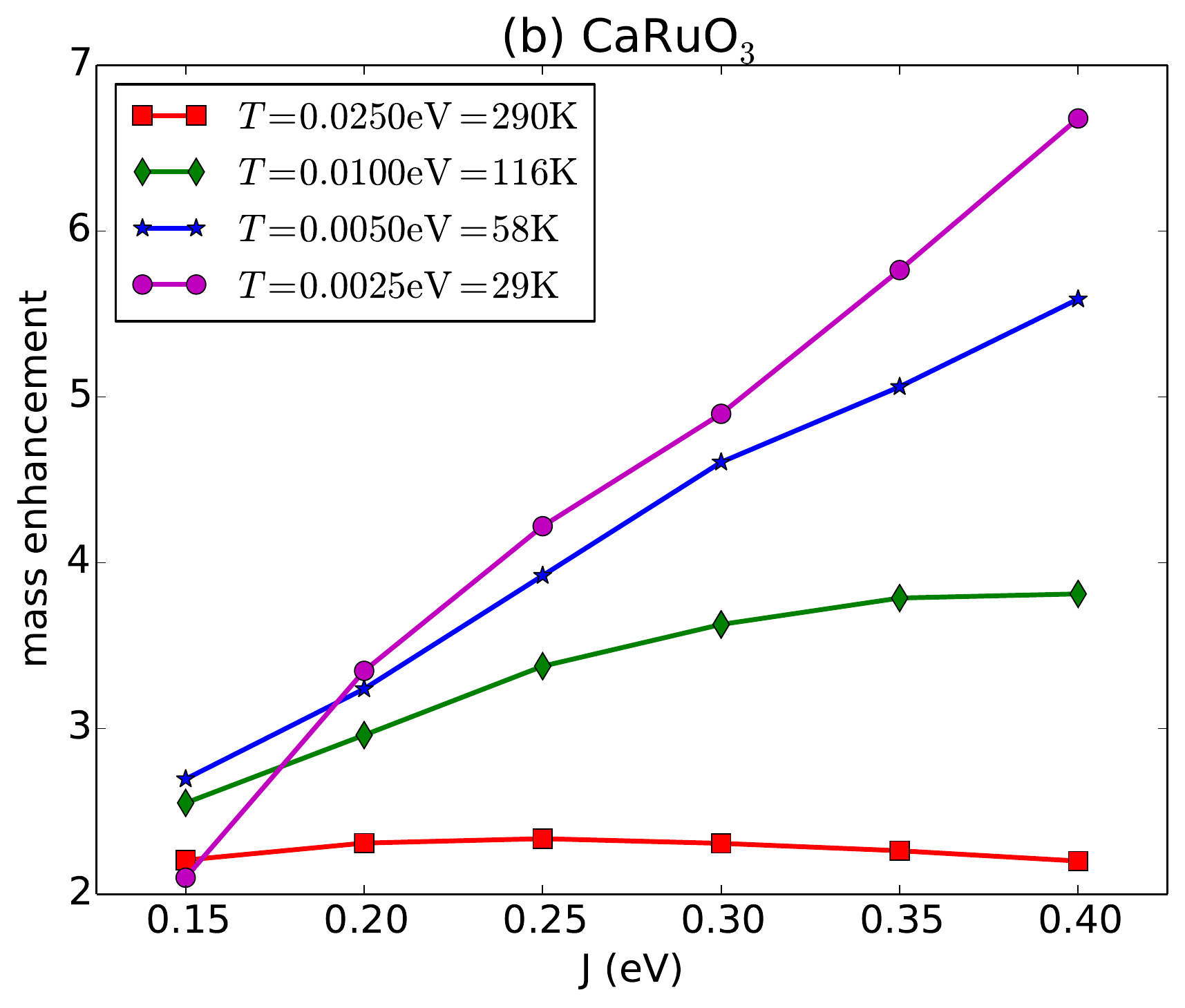}
    \caption{\label{fig:mass_enhancement_J}(Color online) The mass enhancement of (a): SrRuO$_3$ and (b): CaRuO$_3$ vs. the Hund's coupling $J$  calculated at $U=2.3$\,eV  for several different temperatures. For SrRuO$_3$, the larger square indicates  the $J$ value closest to the onset of ferromagnetism at the corresponding temperature. }
\end{figure}

We now use the experimentally measured low temperature specific heat coefficient  to help constrain the parameter values. The measured low-$T$ specific heat of \cro is $\approx 74$\,mJ/molK$^2$ \cite{Cao97}, implying  a mass enhancement $m^\star/m\approx 7$  with respect to the DFT value.  We note that the experimental value has a contribution (of unknown magnitude) from the electron-phonon interaction, so should be regarded as an upper bound on the electronic contribution to the mass enhancement. Figure~\ref{fig:mass_enhancement} shows that $Z^{-1}$ has a marked dependence on $U$ while Fig.~\ref{fig:mass_enhancement_J} shows that $Z^{-1}$ depends even more strongly on Hund's coupling $J$.  As the interaction parameters are not likely to change significantly between the two compounds, we assume that \sro and \cro are described by the same $(U, J)$ values.  Requiring that  the calculated mass enhancement for \cro is close to but not higher than the measured mass, and at the same time  that $J$ be such that SrRuO$_3$ is ferromagnetic and CaRuO$_3$ is paramagnetic allows us to locate the materials on the phase diagram.

At $U=3$\,eV the phase diagram of  Fig.~\ref{fig:phase_diagram} requires that $0.3~\mathrm{eV}\lesssim J\lesssim0.4$\,eV while for $J$ in this range the masses resulting from our calculations are clearly above the experimental value ($\approx 10$  at $T=50$\,K and  clearly increasing as $T$ is decreased; see also the lower panel of Fig.~\ref{fig:5bands}). Thus we argue that the combination of the mass and phase diagram are inconsistent with the possibility that the perovskite ruthenates are in the Mott-dominated regime.  

On the other hand, our results at smaller $U$ indicate that the Hund's metal regime can provide a good description of the basic physics. A relatively wide range of $J$ is found for which \sro is magnetic and \cro is not, while the sensitive dependence of the mass enhancement on $J$, with masses ranging from much smaller than, to rather larger than, the measured value, means that a reasonable $J\sim 0.3$\,eV and $U\sim 2-2.5$\,eV (close to those found from constrained DFT and cRPA methods \cite{Pchelkina07,Mravlje11,Vaugier12}) will account for the basic physics. 

\begin{figure}
 \includegraphics[width=0.95\columnwidth]{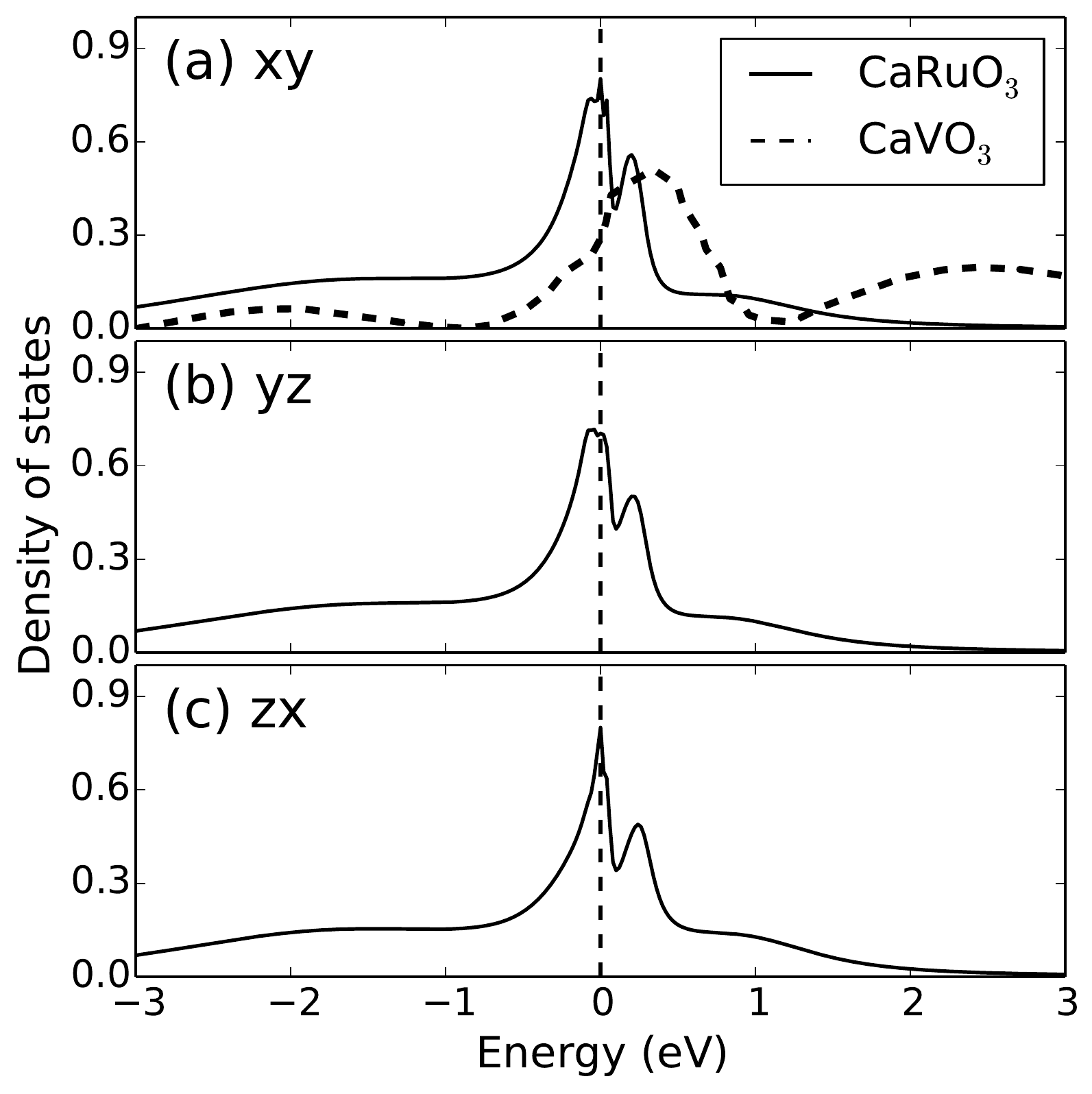}
 \caption{Predicted orbitally-resolved spectral function of CaRuO$_3$. We used $U=2.3$~eV, $J=0.35$~eV, and temperature $T=2.5~\mathrm{meV}\approx29~\mathrm{K}$. Dashed line is the orbitally-averaged spectral function for CaVO$_3$ with data extracted from Ref.~\onlinecite{Nekrasov05}. \label{fig:dmft_dos}}
\end{figure} 

It is interesting to look also at the correlated DOS. We obtained the spectra at real frequencies using the maximum entropy method  \cite{JarrellMaxEnt,Comanac07} to continue the self-energies \cite{Wang09a}  and then using the continued self-energies to construct the orbitally-resolved spectra. The results are presented in  Fig.~\ref{fig:dmft_dos} (solid lines). The densities of states of the three orbitals are quite similar. Comparison of the near-Fermi-level fine structure in the DOS to that shown for the noninteracting model in  Fig.~\ref{fig:dft_mlwf_dos} reveals  that the peaks seen at $\sim -0.4$~eV and $\sim+0.35$~eV in the $xy$ and $yz$ bands of the CaRuO$_3$  DFT DOS  are renormalized differently. The negative frequency peak is pulled up much closer to the Fermi level, appearing at  about $-0.05$~eV in the correlated band structure. The positive frequency peak in the correlated band structure occurs at approximately the  $+0.2$~eV energy of the DFT band structure, thus is renormalized by a much smaller amount. The difference is a manifestation of a large particle-hole asymmetry in the self-energy, whose influence on the spectroscopy was discussed in \cite{Stricker14}. The effect is a fingerprint of Hund's metal physics, as recently noted in Ref.~\onlinecite{wadati14}. See also a very recent preprint \cite{kim_15} for calculations of DOS in SrRuO$_3$ within an approach similar to ours.

Figure~\ref{fig:dmft_dos} also presents for comparison the many-body density of states calculated for the $d^1$ material CaVO$_3$ \cite{Nekrasov05}, in which the correlations are believed to arise from Mott physics (Hund's metal physics requires a higher occupancy of the $d$ level). The spectrum of the Mott material consists of a central peak corresponding to a  uniformly renormalized DFT DOS and clearly visible Hubbard satellites at $-1.5$ and $+3.5$~eV. In contrast the spectrum of the  Hund's material CaRuO$_3$ lacks clear Hubbard sidebands, but as noted above does exhibit a more strongly renormalized and more highly structured quasiparticle peak.  Although the Hubbard bands are more visible in the Mott material, the correlations are in an important sense weaker, with the mass enhancement of CaVO$_3$ only about 2-3 \cite{Nekrasov05} in contrast to the $\sim 7$ that we find  for CaRuO$_3$.

\section{Magnetic phase of $\mathbf{SrRuO_3}$\label{sec:magnetic_sro}}

\begin{figure}[t]
    \centering
    \includegraphics[width=0.9\columnwidth]{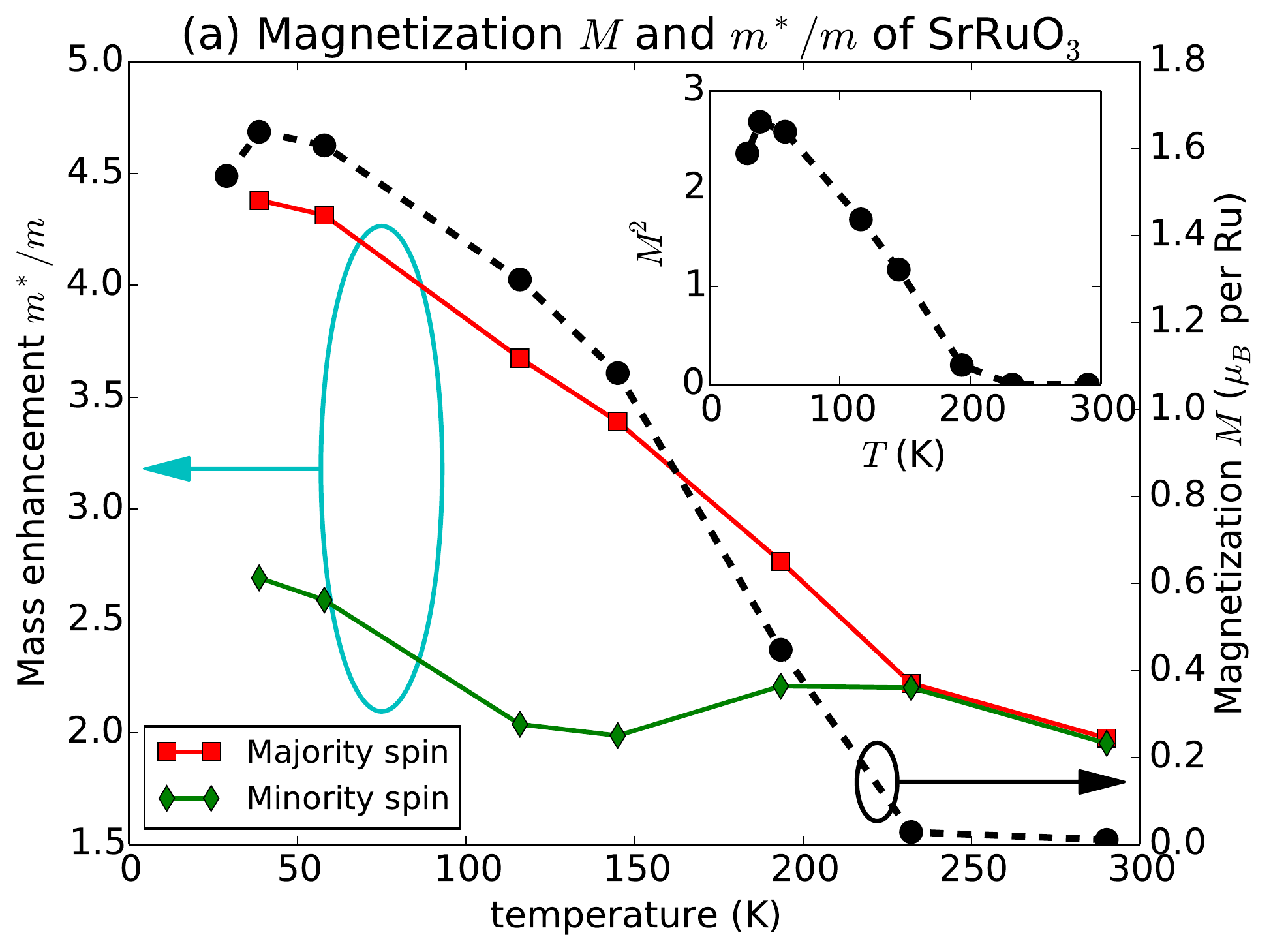}
    \includegraphics[width=0.8\columnwidth]{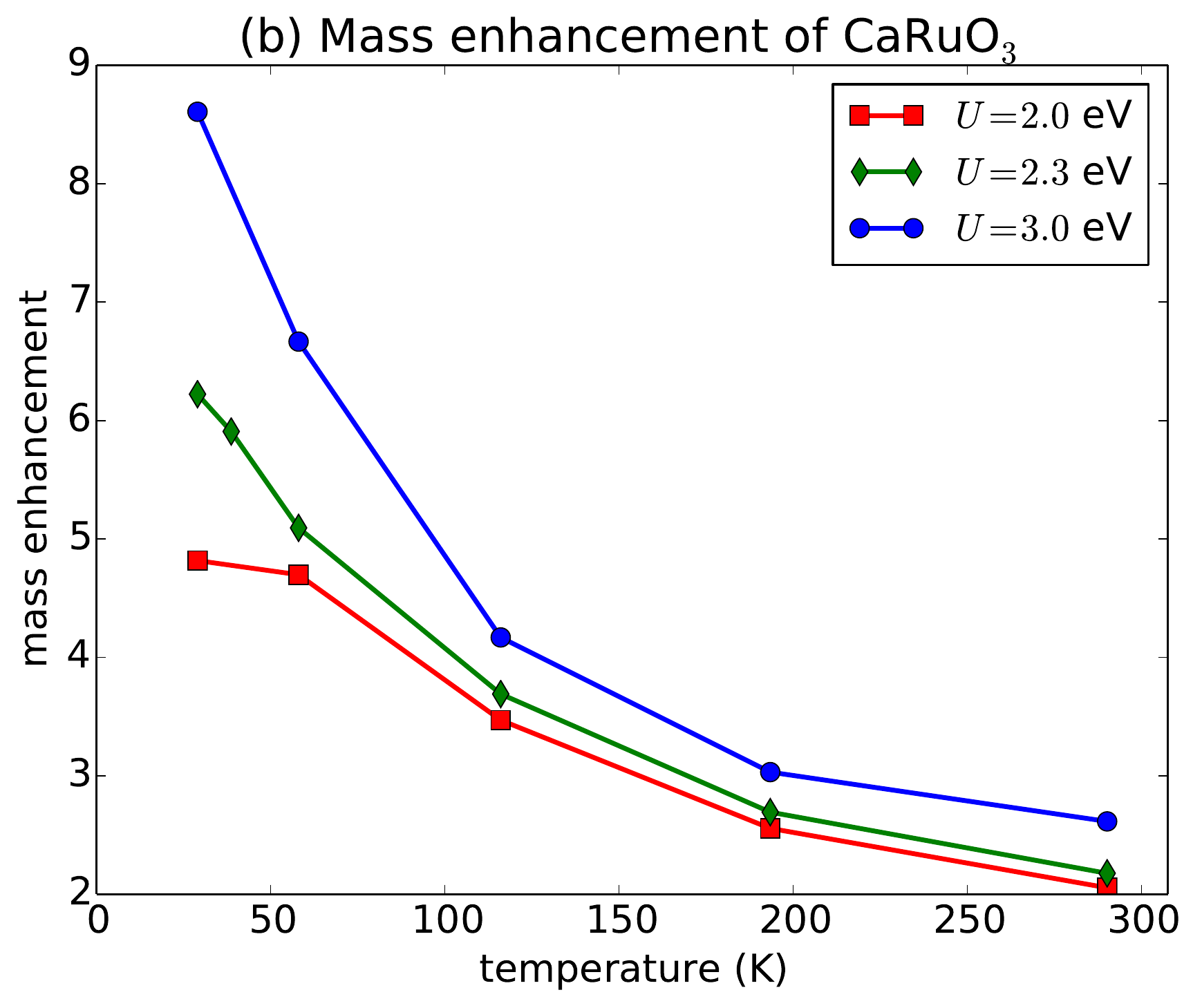}
    \caption{\label{fig:5bands}(Color online) (a) Solid curves: temperature dependence of orbitally-averaged  spin-resolved mass enhancement  of  SrRuO$_3$  calculated at $U=2.3$\,eV and $J=0.35$\,eV (left $y$ axis). Dashed curve: magnetic moment  $M=\langle m_z\rangle$ calculated for same parameters (right $y$ axis).  Inset: temperature dependence of squared magnetic moment $M^2=\left(\langle m_z\rangle\right)^2$. (b) Orbitally-averaged mass enhancement vs. $T$ for CaRuO$_3$ calculated at $J=0.35$\,eV and different $U$ values; note  \cro  is paramagnetic at all $T$.}
\end{figure}

In this section we show that the correlated Hund's metal picture provides an adequate description of the  magnetic phase of \sro, preserving the successful description of the magnetic properties obtained from   DFT calculations \cite{Maiti05,Etz12} while simultaneously providing a good account of the quasiparticle mass. While  a restriction of the correlated subspace to the Ru $t_{2g}$ orbitals is adequate for most purposes, calculations within this scheme yield half-metals, with magnetic moments that saturate at $2\,\mu_B$ per Ru site and with too-small values of the specific heat coefficient. We therefore treat a five-band model that includes both $e_g$ and $t_{2g}$-symmetry orbitals, with the $e_g$ states and $e_g$-$t_{2g}$ interactions treated in a mean field approximation. [The intra-$t_{2g}$ interactions remain as specified in Eq.~\eqref{eq:onsite_SlaterKanamori}]. Inclusion of the $e_g$ orbitals, which lie well above the Fermi level and play no important role in the calculation of dynamical quantities in the paramagnetic phase,  acts to stabilize the magnetization at partial polarization.

The two panels of Fig.~\ref{fig:5bands} compare the mass enhancements of \sro and \cro computed within the five-band model. The lower panel  shows the $T$-dependence of the \cro mass enhancements for three $U$ values. The behavior is in agreement with the three-band computations discussed above: the mass increases as $T$ is decreased, and below a $U$-dependent scale saturates at a $U$-dependent value. At the $J=0.35$\,eV in this figure, we see that the mass is too small at $U=2.0$\,eV and extrapolates to a too-large value at $U=3$\,eV, suggesting that a $U\approx 2.3$\,eV provides a reasonable description of the physics. 

The upper panel shows the temperature dependence of the spin-resolved mass  enhancement of \sro for the parameters $U=2.3$, $J=0.35$\,eV that provide a good description of \cro, along with the temperature dependence of the calculated magnetic moment.   The moment saturates to a value $\approx 1.64\,\mu_B$.  Experiments report values of the magnetic ranging from $0.8$ to $1.6\,\mu_B$ per Ru site \cite{Longo68,Kanbayasi76,Cao97}, with more recent experiments converging on values between $1.4$ and $1.6\,\mu_B$ \cite{Bushmeleva06,Cheng13}.  DFT calculations \cite{Santi97,Singh96,Mazin97,Allen96,Granas14} report magnetic moments consistent with experiment. We see that the Hund's metal picture provides a similarly good level of agreement. The inset reveals that the magnetization has a  mean-field-like temperature dependence $M^2$ linear in $T_c-T$ near the transition; extrapolation to $M^2=0$ indicates a Curie temperature $\approx 200$\,K slightly higher than the experimental $T_c\approx 160$\,K. Fluctuation effects in a three dimensional magnet typically reduce the transition temperature by $\sim 30\%$ relative to the DMFT value (see, e.g. Refs.~\onlinecite{Calderon98,Chattopadhyay00})  so this value also  is very reasonable. 

At temperatures above the magnetic phase transition the mass enhancement of \sro is very close to that of \cro. (The differences discussed above between the paramagnetic phase mass enhancements of the two materials become manifest only at low $T$ where the mass enhancement is large and very sensitive to the spin dynamics controlled by $J$ and the density of states). As $T$ is decreased below $T_c$ we see that the mass enhancement in \sro becomes spin dependent, taking different values for the majority and minority spin channels. The smaller value of the mass in the minority-spin channel may be understood as a phase space effect. The dominant part of the interaction  is between opposite spin species, embodied in the $S^+_m S^-_{m'}$ part of the local interaction. Because we are dealing with a greater than half-filled band, the phase space available for a minority spin electron to scatter into a majority spin state is less than the phase space available for a majority-spin electron to scatter into a minority spin state. More importantly, as the amplitude of the magnetic moment increases we see that the increase in mass is cut off, so the concavity of the $m^\star/m(T)$ curve changes and the $T$ dependence of the mass saturates below $T\sim 50$\,K.  This behavior is a natural consequence of the Hund's metal physics, in which the large mass enhancement arises from slow fluctuations of  spontaneously generated local moments whose formation and dynamics is very sensitive to $J$ and $T$ \cite{Werner08,Georges13}. The quenching of these moments in the ordered phase then cuts off the increase of the  mass enhancement.  The values obtained for the majority-spin mass are in reasonable correspondence with experiment \cite{Allen96} although the contribution of the minority spin channel to the overall specific heat requires further investigation.

\begin{figure}[t]
    \centering
    \includegraphics[width=\columnwidth]{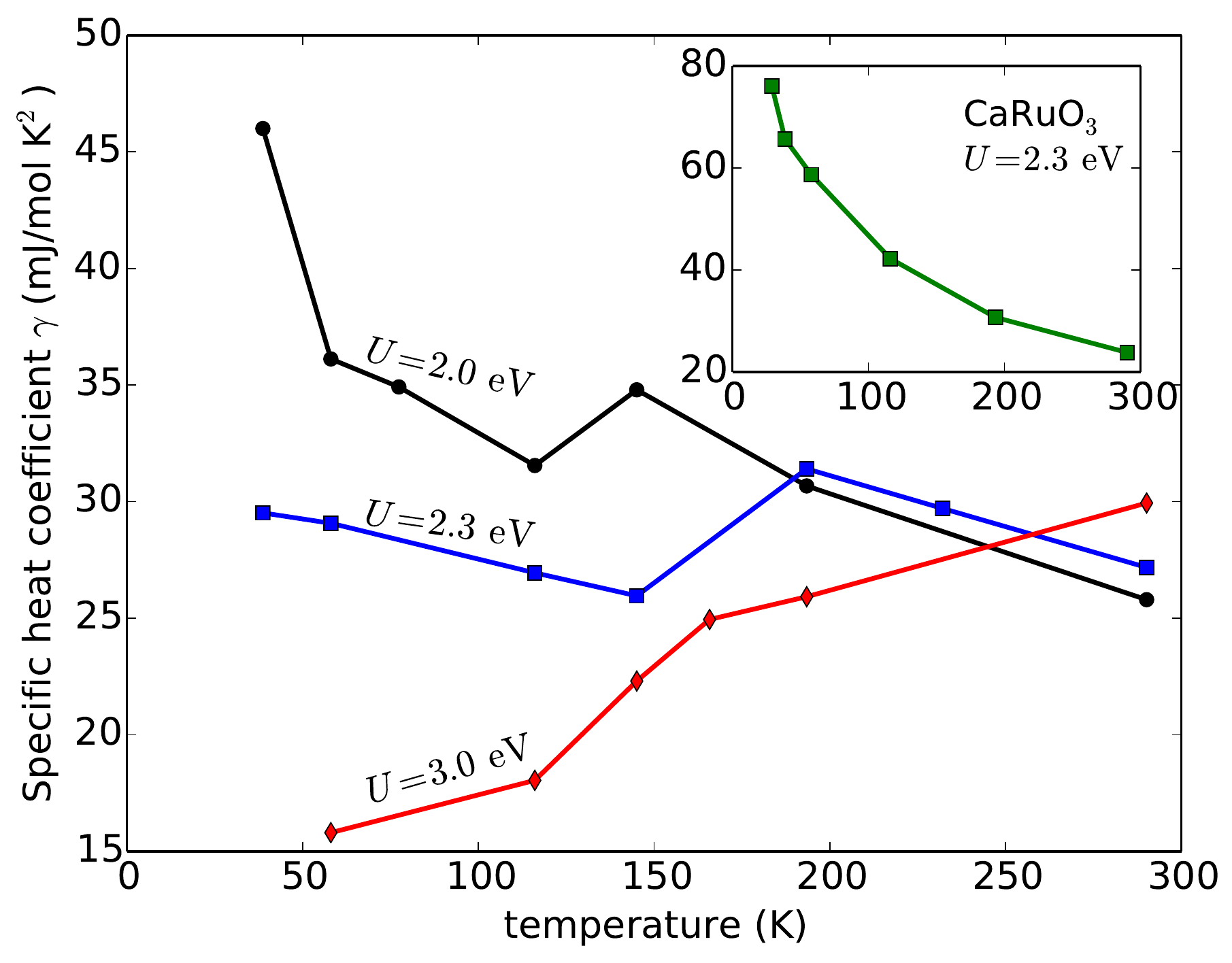}
    \caption{\label{fig:specific_heat}(Color online) Specific heat coefficient $\gamma$ (Sommerfeld constant) calculated for SrRuO$_3$ as a function of temperature $T$ at $J=0.35$\,eV and  $U$ values indicated. Inset: $\gamma$ for CaRuO$_3$ at $U=2.3$\,eV and $J=0.35$\,eV.}
\end{figure}

In Fig.~\ref{fig:specific_heat}, we present an estimate for the specific heat coefficient $\gamma=\lim_{T\rightarrow 0}C/T$  for both materials. We obtain the specific heat by using the quasiparticle approximation $\mathbf{\Sigma}_\sigma(\omega) \approx \mathbf{\Sigma}^0_\sigma + (1-\mathbf{Z}_\sigma^{-1})\omega$  in the formula
\begin{equation}\label{eq:general_form_gamma}
 \gamma = \dfrac{\pi^2k_B^2}{3} \sum_\sigma 
          \dfrac{1}{\pi}\int \dfrac{d^3\vec{k}}{(2\pi)^3}\mathrm{Tr}
          \left[\mathrm{Im}\mathbf{G_\sigma}(\vec{k},i0^-)\mathbf{Z}_\sigma^{-1}\right],
\end{equation}
which is obtained from the quasiparticle approximation to the free energy $F=-\mathrm{Tr} \ln \mathbf{G}$. Here the spin-dependent quantities $\mathbf{\Sigma}_\sigma,\mathbf{G}_\sigma$ and $\mathbf{Z}_\sigma^{-1}$ are $12\times 12$ matrices, accounting for the twelve bands (three for each of the four Ru atoms in the unit cell.) Use of the quasiparticle approximation in Eq.~\eqref{eq:general_form_gamma} is strictly correct only in the low-$T$ Fermi liquid regime, but is expected to be of the correct order of magnitude even at higher temperatures. In the orbital basis that minimizes the hybridization function (see Sec.~\ref{sec:formalism}),  $\mathbf{Z}_\sigma$ is diagonal in site and orbital indices while the four Ru atoms are equivalent up to a rotation in orbital space. In this case, Eq.~\eqref{eq:general_form_gamma} simplifies to
\begin{equation}\label{eq:simplified_form_gamma}
 \gamma = \dfrac{\pi^2k_B^2}{3} \sum_{\sigma m}\dfrac{\mathrm{Im}G^{mm}_\sigma(\vec{r}=0,i0^-)}{\pi}
 Z_{\sigma m}^{-1},
\end{equation}
where the sum is over spin and the $d$ orbitals of one particular Ru atom. Figure~\ref{fig:specific_heat} is produced using Eq.~\eqref{eq:simplified_form_gamma} with $Z_{\sigma m}$ and $\Sigma^0_{\sigma m} = \mathrm{Re}\Sigma_{\sigma m}(i\omega_n\to i0^+)$ obtained from DMFT.

We see  from Fig.~\ref{fig:specific_heat} that the value $U=2.3$\,eV selected on the basis of the phase diagram and \cro mass reproduces well the  low-T experimental results   ($\gamma\sim 30$~mJ/mol\,K$^2$ for SrRuO$_3$ and $75$~mJ/mol\,K$^2$ for CaRuO$_3$ \cite{Cao97,Allen96}). Larger (3.0\,eV) or smaller (2.0\,eV) value of $U$ are inconsistent with experiment.  Essential in obtaining this level of agreement is obtaining a correct estimate for the magnitude of the saturation moment. A too small moment ($U=2.0$\,eV) leaves active spin fluctuations which cause a further increase in the mass, while if the moment is too near to saturation ($U=3.0$\,eV) the   $\gamma$ is too small.

\section{Conclusions\label{sec:conclusions}}

In this work, we have investigated the interplay of electronic correlations and lattice distortions in  the perovskite ruthenates SrRuO$_3$ and CaRuO$_3$ using a density functional treatment of the basic electronic states and treating correlations via dynamical mean-field theory with a CT-HYB impurity solver \cite{Werner06}. We determined the  behavior of a general class of models motivated by the physics of the ruthenates, finding that  ferromagnetism depends on (1) how far materials are from Mott insulating phase and (2) positions of DOS peaks with respect to the Fermi level. The latter factor is more important for small and intermediate correlations, while the former  controls the behavior at strong correlation.

Our main results are presented in the phase diagram shown in  Fig.~\ref{fig:phase_diagram}.  The choice of $U=2.3$\,eV value (far from critical value for the Mott insulating phase) and $J=0.35$\,eV gives a calculated mass enhancement for CaRuO$_3$ in reasonable agreement with  experiment.  The mass enhancement is a sensitive function of the distance of the material from the ferromagnetic phase boundary. Several experimental works propose that very weak disorder may induce spin-glass behavior in CaRuO$_3$ \cite{Cao97,Felner00,Mukuda99}, suggesting that the material is very close to a magnetic phase boundary, consistent with our results. The same calculations predict that SrRuO$_3$ becomes ferromagnetically ordered below a Curie temperature somewhat less than room temperature. The onset of ferromagnetism cuts off the low-$T$ increase of the mass, and the calculated ferromagnetic-state value  is found to be much smaller than that of paramagnetic CaRuO$_3$ and also in good agreement with experiment. We therefore concluded that the ruthenates are far from the Mott insulating phase and may be identified as Hund's metals.  

A test of this picture is that the predicted density-of-states for CaRuO3 is typical of a Hund's metal: it shows only weak signatures of Hubbard bands despite the large renormalizations at low frequency; in contrast, the density of states of a material located near the Mott critical point would exhibit a much larger redistribution of weight away from the Fermi level and would display clear Hubbard sidebands.   Closely related to this point is the local susceptibility shown in the inset to Fig.~\ref{fig:mass_enhancement} which is very strongly enhanced relative to the band theory value, demonstrating that the large renormalizations come from strongly enhanced local spin fluctuations.  A further consequence of the identification of the materials as Hund's metals is that the reason for the occurrence of  ferromagnetism  in SrRuO$_3$ and its absence in CaRuO$_3$ is the difference in density of states of the two materials.

Our study has certain limitations. First, it employs the single-site dynamical mean field approximation. While this captures many important aspects of local energetics and material trends, it is not necessarily quantitatively accurate for $d=3$ dimensional materials. In particular, the DMFT theory includes dynamical effects of momentum averaged spin fluctuations but does not include all of the effect of quantum critical spin fluctuations which are important near magnetic transitions \cite{book:Moriya85}.   Spin fluctuations are known to have particularly important consequences in quasi-two-dimensional materials and  have been extensively discussed in the context of the two-dimensional ruthenate material Sr$_2$RuO$_4$; their quantitative importance in $d=3$ is less \cite{book:Moriya85} but the actual contribution to physical properties, both in the  ruthenates, and more generally  in the context of  DMFT,  is an important open question which warrants further research.

Second, we have used the ``frontier orbital'' approximation in which correlations are applied to the $d$-derived near-Fermi-surface bands. However, the good qualitative and even semiquantitative agreement between our calculations and experiment  justifies this approximation \textit{a posteriori}. Our results thus unambiguously indicate that the perovskite ruthenates are in the Hund's metal class of materials, with strong correlation effects driven by the $J$ rather than the $U$ term of the interaction. This finding resolves the tension between the successful DFT account of the magnetism and the thermodynamic, transport and optical results indicating strong correlations. 

Third, we have neglected spin-orbit coupling which splits the sixfold degenerate $t_{2g}$ manifold into fourfold and twofold degenerate manifolds and tends to reduce bandwidths, and also can lead to spin anisotropy \cite{koster12rmp} and to interesting entanglement effects e.g. on superconducting wave functions \cite{veenstra_14}. If the  spin-orbit induced splitting (as renormalized by interactions) is large enough to significantly change the structure of the $t_{2g}$ manifold, for example by fully separating the twofold- and fourfold-degenerate submanifolds, then the basic picture presented here would not apply. On the other hand, if the spin-orbit coupling is not too large, then the tendency of $J$ to favor high-spin states in which electrons are more or less equally distributed  over orbitals will tend to suppress spin-orbit effects. The crossover as spin-orbit coupling is increased at fixed $J$ will have something of the character of a high-spin/low-spin transition and is thus expected to be abrupt. DFT calculations \cite{Havercourt08} for related material (Sr$_2$RuO$_4$) yield a spin-orbit  splitting of at most $\sim 200$~meV and with no significant alteration of the basic properties of the bands. This, and the fact that calculations of the kind presented here reproduce well the properties of both pseudocubic and layered ruthenates  (see Refs.~\onlinecite{Mravlje11,Stricker14,kim_15}) suggests that in fact spin-orbit coupling does not have a strong effect on the basic electronic properties of the ruthenates. The issue however warrants more thorough investigation.

We conclude by indicating few directions for the future work. Our calculations show an intricate interplay between lattice structure and correlation effects, mediated by the lattice-induced changes in the near-Fermi-surface density of states. Films show a different pattern of rotation and tilt than do bulk materials, and in films the rotation angles may be manipulated by strain. Calculations of the strain dependence of the mass enhancement and magnetic moment in thin films of \sro and \cro will be very interesting to perform and compare to experiment~\cite{Herklotz2014,Tripathi2014} and to previous LSDA results~\cite{zayak06,zayak08}. Studying theoretically the ruthenates within a wide energy window to include oxygen $p$ and $e_g$ bands, which are both close to the frontier $t_{2g}$ bands, is an interesting subject, too.  Finally, the formalism and physical picture provided here constitute a potentially useful starting point for investigations of impurity-induced magnetism in CaRuO$_3$.

\section*{Acknowledgments}

We thank A. Fujimori and H. Wadati for helpful discussions.  H.T.D. acknowledges support from the Deutsche Forschungsgemeinschaft (DFG) within projects FOR 1807 and RTG 1995, as well as the allocation of computing time at J\"ulich Supercomputing Centre and RWTH Aachen University through JARA-HPC. J.M. acknowledges support of Slovenian research agency under program P1-0044. A.G. acknowledges a grant from the European Research Council (ERC-319286 QMAC) and support from the Swiss National Science Foundation (NCCR-MARVEL). A.J.M. acknowledges support from NSF-DMR-1308236.

\appendix

\section{Wannier Fits \label{Appendix:BandTheory}}
\begin{figure}[t]
\centering
\includegraphics[width=\columnwidth]{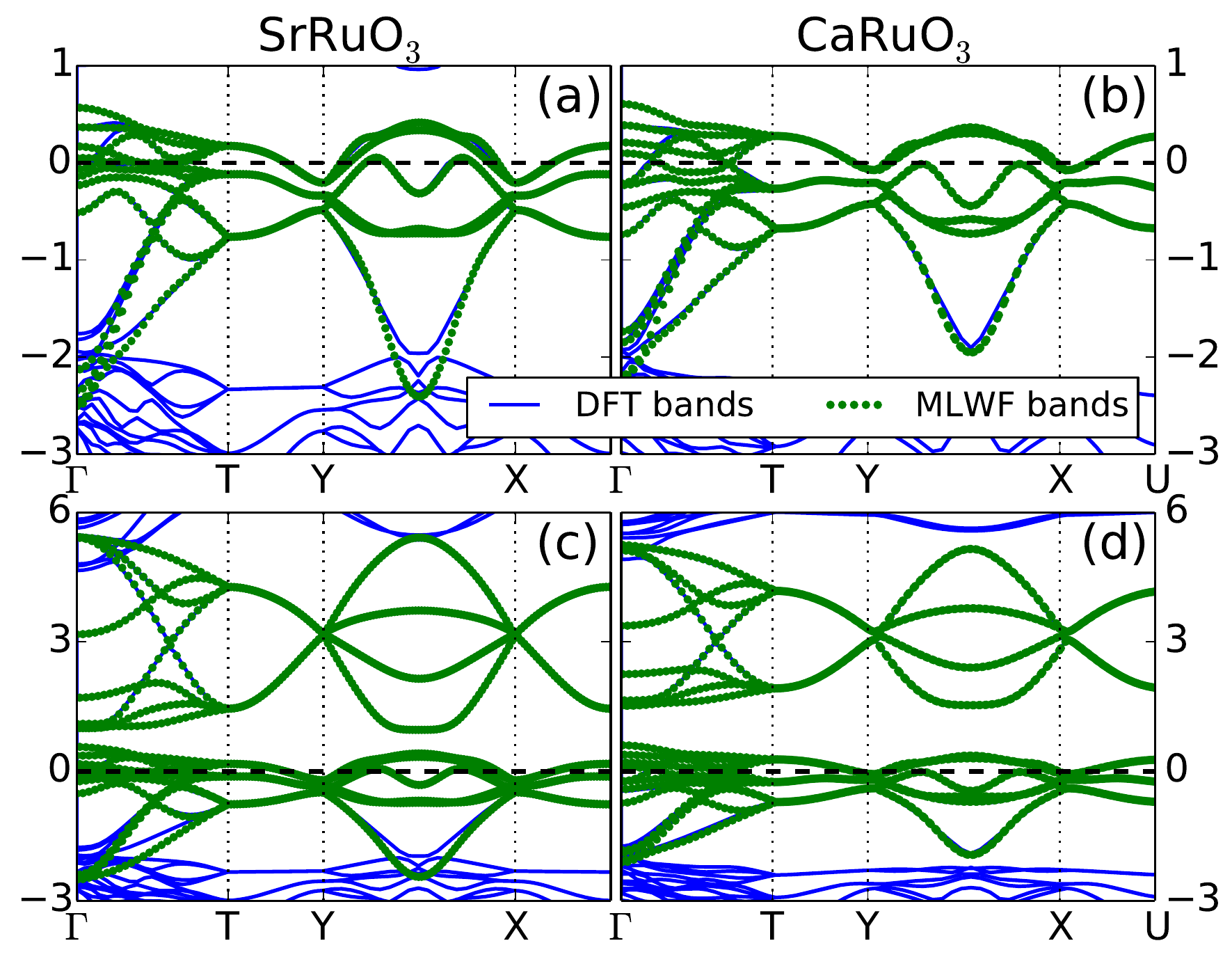}
\caption{\label{fig:dft_mlwf_bands}
(Color online) DFT band structure (solid curves, blue online) and the corresponding MLWF fits (dots, green online) for SrRuO$_3$ (left column) and CaRuO$_3$ (right column) using their corresponding experimental orthorhombic structure \cite{Jones89,Bensch90}. (a) and (b): MLWF fits for $t_{2g}$ subspace. (c) and (d): MLWF fitting for $t_{2g}$-$e_g$ subspace. The band structures are plotted along the $k$ path $\Gamma\to T\to Y\to X\to U$ where $\Gamma=000,T=\pi\pi\pi,Y=\pi\pi0,X=\bar{\pi}\pi0$ and $U=\bar{\pi}\pi\pi$ in the first Brillouin zone of the orthorhombic structure. The horizontal dashed line marks the Fermi level. }\end{figure}

Figure~\ref{fig:dft_mlwf_bands} presents the band dispersion obtained from our density functional calculations in several high symmetry directions for SrRuO$_3$ and CaRuO$_3$. The energy window has to be adjusted  to capture all the states belong to the subspace. For the $t_{2g}$-only subspace [Fig.~\ref{fig:dft_mlwf_bands}(a) and \ref{fig:dft_mlwf_bands}(b)], an energy window from $-3$ to $1$\,eV is used. For the $t_{2g}$-$e_g$ subspace [Fig.~\ref{fig:dft_mlwf_bands}(c) and \ref{fig:dft_mlwf_bands}(d)], the energy window ranges from $-3$ to $6$\,eV.

The frontier orbitals are seen to be well separated from other bands at higher energy, and as a result the MLWF fits are adequate over most of the relevant energy range. However, at the bottom of the $t_{2g}$ bands, some overlap with the  oxygen $p$ bands occurs, especially in the case of SrRuO$_3$. The band overlap occurs near the $\Gamma$ (zone center) point of the Brillouin zone, at which orbital characters are well defined and $p$-$d$ hybridization is minimal. We have verified for the density of states projected to atomic orbitals of SrRuO$_3$ (not shown), the Ru $d$ character is nonzero until $\sim -2.5$\,eV, indicating that the Wannier fitting is thus reasonable even in this region of band overlap. 

As noted in the main text, if only the $t_{2g}$ manifold is included, the Wannier functions produced by the {\sc wannier90} code \cite{Mostofi08} are  aligned with the local octahedral axes and the DMFT hybridization function can be constructed directly from the projection of the Kohn-Sham Hamiltonian onto the Wannier basis. If however all five $d$ orbitals are included (as in our magnetic calculations) the orbitals produced by the the {\sc wannier90} code are not properly aligned to the local symmetry axes, and must be rotated, in order to minimize the off-diagonal terms in the DMFT hybridization functions. We find that the desired rotation is the one that diagonalizes the site-local terms in the projection of the Kohn-Sham Hamiltonian onto the Wannier basis. 

Our convention for the orbitals whose self-energy is shown in Fig.~\ref{fig:coherence} is as follows. The $Pnma$ structure has three lattice vectors conventionally denoted $\vec{a}$, $\vec{b}$ and $\vec{c}$. The orbitals are labeled in terms of  pseudocubic $\hat{x}$, $\hat{y}$, $\hat{z}$ directions defined as those closest to 
\begin{eqnarray}
\vec{a}&=& \hat{x}+\hat{y}, \\
\vec{b}&=& 2\hat{z},\\
\vec{c}&=& \hat{x}-\hat{y},
\end{eqnarray}
and Fig.~\ref{fig:coherence} presents the diagonal components of the self-energy for the Ru ion at position $(0,0,0.5)$ in the system defined by the $Pnma$ $\vec{a}$, $\vec{b}$, and $\vec{c}$ lattice vectors

\section{Criteria for determining the electronic phases and phase boundaries}
\label{sec:criteria}

\subsection{Ferromagnetic-paramagnetic phase boundary}

To locate the ferromagnetic-paramagnetic phase boundary we follow our previous work \cite{Dang13} and  compute the inverse magnetic susceptibility as $\chi^{-1}=h/m$ with $m$ the calculated magnetization and $h$ an applied field chosen to be small enough that the $m(h)$ curve is linear (typically $0.01$\,eV, but can be smaller at low temperatures). We perform the calculation at several temperatures,  fit the result to a straight line $\chi^{-1}(T)=A(T-T_c)$ and determine the phase according to whether the $T=0$ extrapolation $\chi^{-1}(T=0)=-AT_c$ is positive or negative. 

\begin{figure}[t]
    \centering
    \includegraphics[width=0.7\columnwidth]{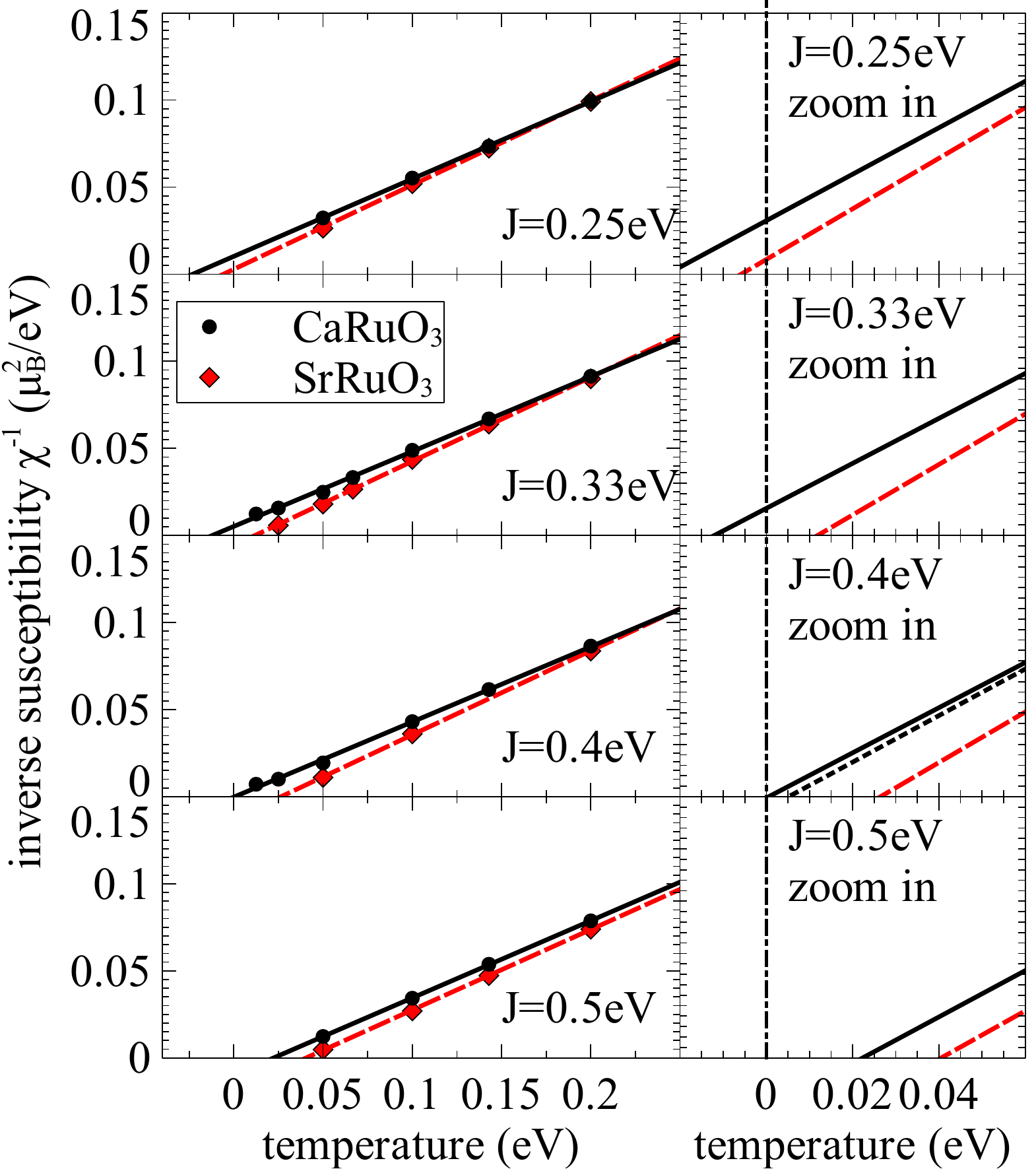}
    \caption{(Color online) Left column: the evolution of the Curie temperature for CaRuO$_3$ and SrRuO$_3$ with $J$ increased from $0.25$\,eV to $0.5$\,eV and $U=3$\,eV kept fixed; the fitting lines (solid for CaRuO$_3$ and dashed lines for SrRuO$_3$) and all data points are included. Right column: expanded plots for smaller range of temperature near $0$. The dotted line at $J=0.4$\,eV is the fitting line for CaRuO$_3$ if data of two lowest temperatures are neglected. The vertical dashed line marks  zero temperature.}
    \label{fig:Tc_evolve}
\end{figure}

For each point on the $U$-$J$ phase diagram, we use at least four different temperatures, typically $T=0.2, 0.14, 0.1$ and $0.05$\,eV,  to determine the Curie temperature (for some points close to the phase boundary, we go to lower temperatures). Within the magnetic phase the  inverse susceptibility $\chi^{-1}$ is typically linear in this range of temperature, thus $T_c$ is easily obtained. The phase boundary is then specified by linear interpolation between the points of lowest positive and largest negative $T_c$. 

Figure~\ref{fig:Tc_evolve} demonstrates the approach presenting $\chi^{-1}(T)$  for different $J$ values at a fixed $U=3$\,eV. We see clearly that \sro has a greater tendency to ferromagnetism than \cro, with the difference being more pronounced at higher $J$.  We note, however, within the paramagnetic phase or in the magnetic phase very close to the phase boundary, $\chi^{-1}$ starts to bend away from the high-$T$ linear extrapolation at low temperature, as can be seen from a close examination of the $\chi^{-1}$ for \cro for $J=0.33$ and $0.4$\,eV. For these data we have pushed the \cro calculation to the lower temperatures $T=0.025$ and $0.0125$\,eV.  The bending away from the Curie-Weiss curve  is a signature of the onset of Fermi-liquid coherence and leads to uncertainty in specifying the magnetic phase boundary. For example, at $J=0.4$\,eV, the linear extrapolation gives $T_c=0.0006$\,eV if the points at  $T=0.025$ and $0.0125$\,eV are included (solid line), but if these two temperatures are excluded (as in most of our calculations for building the phase diagram), we would obtain $T_c=0.0052$\,eV (the dotted line in ``zoom in'' panel of Fig.~\ref{fig:Tc_evolve}). Therefore we expect the error bar for $T_c$ of about $0.0045$\,eV. Our $U$-$J$ phase boundary contains similar uncertainties, but determining  the precise error bars  $\Delta U, \Delta J$ requires heavy calculations to go to ultra-low temperatures. The uncertainties arising from the onset of coherence do not affect the qualitative conclusions of this paper.

\subsection{Metal-insulator phase boundary}

We define whether the material is  insulating or metallic according to whether or not the electron spectral function (many body density of states)  $A(\omega)$ vanishes at the Fermi level $\omega=0$ as $T\rightarrow 0$. We determine $A(\omega=0)$ from the imaginary part of the measured Matsubara Green's function $G(i\omega_n)$ such that $\mathrm{Im}G(i\omega_n)\to -\pi A(\omega=0)$ when $\omega_n\to 0$. In practice, we observe $\mathrm{Im}G(i\omega_n)$ at several lowest Matsubara frequencies: if it bends towards zero, the state is insulator, whereas it goes away from zero, it is metallic. By fixing $J$ and gradually increasing $U$, the critical value $U_c$ is determined if the $\mathrm{Im}G(i\omega_n)$ bending changes at low frequencies. (See  \cite{Dang14} for details). 

In single-site dynamical mean-field theory, the metal-insulator phase boundary has a complicated structure at low $T$, with a line of first order transitions emerging  $T=0$ and there is second order transition at interaction values $U_{c2},J_{c2}$ and terminating at a critical endpoint $U_{c1},J_{c1}$, with $U_{c1}$ typically $0.8-0.9U_{c2}$, there exists a temperature  $T_{MIT}$  where $U_{c1}(T_{MIT}) = U_{c2}(T_{MIT})$. Above  this temperature  there is only a crossover from metallic to insulating state \cite{Georges96}. We start all of our calculations from a metallic initial condition and the true  metal-insulator transition is at a slightly lower temperature than the $T_{MIT}$ we find,  and the $U_c$ we determine  is closer to $U_{c2}$ than $U_{c1}$.

\bibliography{scro}
\end{document}